\documentclass[12pt,aps,twocolumn,pra, superscriptaddress,reprint,floatfix]{revtex4-2}
\usepackage{graphicx}
\usepackage{setspace}
\usepackage{color}
\usepackage{tabularx}
\usepackage{multirow}
\usepackage{tabularx}
\usepackage{rotating}

\usepackage{booktabs}
\usepackage{hyperref}
\usepackage{lipsum}
\usepackage{siunitx}
\usepackage{bm}
\usepackage{mwe}
\usepackage{amsmath,amssymb}
\newsavebox{\measurebox}

 \newcommand{\bea}{\begin{eqnarray}}
	\newcommand{\eea}{\end{eqnarray}}
\newcommand{\bes}{\begin{subequations}}
	\newcommand{\ees}{\end{subequations}}
\usepackage{stackengine}

\usepackage{BOONDOX-cal}
\usepackage{rotating}
\usepackage{setspace}
\usepackage{tabularx}
\usepackage{array}
\newcolumntype{Y}{>{\centering\arraybackslash}X}
\usepackage{longtable}

\begin{document}
\bibliographystyle{revtex4-1}
\title{Modulational instability in $\mathcal{PT}$-symmetric Bragg grating structures with four-wave mixing}
\author{I. Inbavalli}
\affiliation{Department of Physics, Presidency College (Autonomous), Chennai - 600 005, India}
\author{K. Tamilselvan}%
\affiliation{Department of Nonlinear Dynamics, School of Physics, Bharathidasan University, Tiruchirappalli - 620 024, India}
\author{A. Govindarajan}
\email{Corresponding author: govin.nld@gmail.com}
\affiliation{Department of Nonlinear Dynamics, School of Physics, Bharathidasan University, Tiruchirappalli - 620 024, India}
\author{T. Alagesan}
\affiliation{Department of Physics, Presidency College (Autonomous), Chennai - 600 005, India}
\author{M. Lakshmanan}
\affiliation{Department of Nonlinear Dynamics, School of Physics, Bharathidasan University, Tiruchirappalli - 620 024, India}
\begin{abstract}

We investigate the dynamics of modulational instability (MI) in $\cal PT$-symmetric fiber Bragg gratings with a phenomenon of intermodulation known as four-wave mixing (FWM).  Although the impact of FWM has already been analyzed in the conventional systems, the inclusion of gain and loss, which induces the notion of $\cal PT$- symmetry, gives rise to many noteworthy outcomes. These include the manifestation of an unusual double-loop structure in the dispersion curve,  which was unprecedented in the context of conventional periodic structures. When it comes to the study of MI, which is usually obtained in the system by imposing a small amount of perturbations on the continuous wave by executing linear stability analysis, different regimes which range from conventional to broken $\cal PT$- symmetry tend to create quite a few types of MI spectra. Among them, we observe a unique MI pattern that mimics a tilted two-conical structure facing opposite to each other. In addition, we also address the impact of other non-trivial system parameters, such as input power, gain and loss and self-phase modulation in two important broad domains, including normal and anomalous dispersion regimes under the three types of $\cal PT$- symmetric conditions in detail.

\end{abstract}
\maketitle
\section{Introduction}
In the context of optics, four-wave mixing (FWM) refers to the third-order nonlinear interaction that specifically emerges when nonlinear interactions occur among four optical waves, generating new optical waves. In practice, the effect of FWM is generally identified as a degrading factor in the performance of many optical communication systems \cite{Kas2010, Giles}. Nevertheless, the FWM interaction contributes to various potential applications, such as wavelength conversion \cite{Morichetti2011, Chen2015},  parametric oscillator \cite{Zhang}, optical switching \cite{Petersen2015}, new frequency comb generation \cite{DelHaye2007}, logic gates \cite{Wang2009, Li2009}, and quantum information processing \cite{Fulconis2007, Ozeki2006}. Furthermore, the notion of FWM contributes to the advancement of various coherent nonlinear processes, including coherent Raman scattering, multi-dimensional spectroscopy \cite{Cundiff2013}, and impulsive stimulated Rayleigh, Brillouin, and Raman scattering \cite{Dhar1994, Bencivenga2014, Tanaka2002, Bencivenga2009}. Owing to its unique properties, the effect of FWM has attracted unprecedented attention in various physical contexts, including single-mode fiber \cite{Chen2011}, highly nonlinear fiber \cite{Liu2008}, polarization-maintaining fiber (PMF) \cite{Guasoni2012}, and fiber Bragg gratings (FBGs) \cite{Bencivenga2015}. Among these, FBGs are notable ones: for instance, the effect of FWM has enhanced the characteristics of FBGs giving rise to an improved design of sensors with a high measuring range \cite{Ghosh} and other technologies as well \cite{Kas2010}.

FBGs are considered as one of the most important integrated devices, offering potential candidates for the next generation of all-optical networks due to their low insertion losses, narrow bandwidths, and high reflectivity. They also provide promising opportunities for practical applications, including wavelength-stabilized pump lasers, dispersion compensators, narrow-band filters, sensors, and add-drop multiplexers \cite{Kas2010, Giles, Hil1997}. As with the FBGs operating in the linear regime, the nonlinearity-enabled FBGs exhibit their own advantages. For example, when an FBG is combined with Kerr nonlinearity, it supports a variety of solitons such as Bragg solitons, gap solitons and so on. The theoretical prediction of such solitons was initially reported in Ref. \cite{Sterke}, and subsequent experimental demonstrations were conducted in FBGs by Eggleton et. al \cite{Eggleton}. This advantage of nonlinear FBGs has stimulated a variety of intriguing phenomena, such as optical switching \cite{switching}, pulse compression \cite{compression}, optical bi- and multi-stability \cite{bistability}, modulational instability \cite{MI} and so on. The notion of nonlinear FBGs (NLFBGs) has been extended to the rapidly developing area of non-Hermitian physics quite recently \cite{Miri}. 

To introduce the quantum mechanical concept of non-Hermiticity in a nonlinear optical system, including NLFBG-like system, one of the feasible ways is to adopt a complex refractive index with an equal amount of gain/loss profile \cite{Miri}. It is found to be an exact equivalent implication of non-Hermitian concept in optics, as the notions of parity ($\cal P$) and time ($\cal T$)-symmetry have been introduced to demonstrate that a non-Hermitian Hamiltonian can admit eigenvalues with real energy spectra in quantum mechanics \cite{Ben1998, Ben2002, Ben2007, Chr2003, Gan2007, Makris2008, Rut2010, Guo2009}. In optics, the concept of $\cal PT$-symmetry serves as a vital source, giving rise to many new and unusual dynamics, such as non-reciprocity \cite{Makris2010}, double refraction \cite{Makris2008} and so on. The advantage of $\cal PT$-symmetry in not only FBGs but also in other optical devices is to produce stable formations of localized  wave structures and thereby exploring their different dynamics under various linear and nonlinear effects \cite{Suchkov}. $\cal PT$-symmetric effects in the nonlinear optics domain has stimulated various striking phenomena, such as unidirectional wave transport at exceptional points \cite{Lin2011}, coherent complete absorption in coupled resonators \cite{sun2014}, nonreciprocal dynamics \cite{Makris2010}, and so forth. Apart from these, the $\cal PT$-symmetric notion has demonstrated many intriguing phenomena that include the coexistence of perfect absorption and lasing \cite{Huang2014}, unidirectional invisibility \cite{Lin2011}, nonreciprocal light propagation dynamics \cite{Kulishov2005}, soliton switching \cite{govindarajan2018}, and novel optical bistable states \cite{raja2019multifaceted, PhysRevA.100.053806} to name to few.

Among the many other nonlinear optical phenomena, the study of modulational instability (MI) is ubiquitous, producing an exponential growth of continuous wave (CW) structures when an infinitesimal noise is imposed on it. It has also been recognized as a precursor for the formation of solitons \cite{Tai, Hasegawa}. Having originated in fluid dynamics, where it is known as the Benjamin-Feir instability \cite{Ben1967}, MI has attracted considerable attention in various areas of physics, including optics \cite{Tai}, solid-state physics \cite{Remoi99}, plasma physics \cite{Akhtar2017}, and electrical lines \cite{Remoi99}. The striking features of MI dynamics have led to many noteworthy features in nonlinear light-matter interactions, such as Fermi-Pasta-Ulam-Tsingu recurrence in optics \cite{Simaeys} and formation of Akhmediev breathers and Peregrine solitons \cite{Erkintalo}, which can be observed during the nonlinear stage of MI in optical fibers \cite{Kraych, Zakharov}. In the literature, the implementation of MI could be utilized to achieve several potential applications, including the generation of terahertz frequency pulses with high repetition rates \cite{Greer}, supercontinuum generation \cite{Dudley}, and the development of optical frequency combs \cite{LeoFC, Del}. These MI characteristics in nonlinear fiber optics have been investigated both analytically and numerically across a wide range of nonlinear media. Furthermore, they have been extended to an emerging area of $\cal PT$-symmetric optics \cite{ZakharovMI, Kivshar}. While MI dynamics in conventional FBGs with different nonlinearities have been extensively studied by many researchers at various times \cite{ Lit2001, Por2005, Por2005AIP}, studies of MI in $\cal PT$-symmetric FBGs are very limited \cite{Sar2014, Tamil}. Motivated by the above advancements of MI, in this article, we attempt to investigate the MI dynamics in both conventional and $\mathcal{PT}$-symmetric FBGs with the modulation of the Kerr nonlinearity.

This article is organized as follows: After a detailed introduction, we present the mathematical derivation for the proposed system in Sec. II. Following this, we provide a procedure for finding the dispersion relation of the governing equations in Sec. III. In Sec. IV, a detailed analysis of the linear stability approach for the governing model is given, where we examine the impact of MI dynamics in the anomalous and normal dispersion regimes by tuning the dispersion parameter. Additionally, we explore the effects of various physical parameters, such as pump power, self-phase modulation (SPM), and FWM, on the MI phenomenon under different $\cal PT$-symmetric conditions in the proposed model. Finally, we summarize the important outcomes in Sec. V.

\section{Model} \label{Sec:II}
In this section, we formulate the fundamental governing equations which are involved in describing the propagation of light in a finite-length periodic fiber Bragg grating structure. This structure incorporates equal amount of gain and loss profiles. Specifically, the distribution of the refractive index $(n(z))$ for the aforementioned system is formulated as follows:
\bea\label{Eq1}
&&n(z)=n_{0}+{\Delta}n_{R}\cos\left(\frac{2\pi z}{d}\right)\pm i\Delta n_{I}\sin\left(\frac{2\pi z}{d}\right)\nonumber\\
&&+n_{2}|E|^{2}+{\Delta}n_{2}\cos\left(\frac{2\pi z}{d}\right)|E|^{2}.
\eea
Here, the parameter $n_{0}$ indicates the average refractive index profile of the medium, and $\Delta n_{R}$ represents the variation in the real part of the refractive index while ${\Delta}n_{I}$ indicates the imaginary part of the index along the grating period of $d$ with a device distance $z$. 
 The last two terms represent the intensity-dependent nonlinear profiles, with $n_2$ denoting the average value of the nonlinear refractive index and $\Delta n_{2}$ representing its modulation profile. The dynamics of the considered system can be expressed through the following time-dependent Helmholtz equation, incorporating the spatial variation $(z)$ of the refractive index as defined in  Eq. \eqref{Eq1} for the optical field $E$,
\bea
\frac{\partial^{2} E}{\partial z^{2}}+\frac{\partial^{2} E}{\partial t^{2}}+k^{2}\frac{n^{2} (z)}{n_{0}^{2}}E=0, \label{Eq2}
\eea
where $t$ refers to the time coordinate and the parameter $k$ represents the wave number. Let us consider an optical field, denoted as $E(z, t)$, comprising forward and backward waves propagating inside the $\cal PT$-symmetric FBGs as
\bea \label{Eq3}
E(z,t)=U(z,t) \exp(i (k z-\omega_{0} t)+\nonumber\\ V(z,t) \exp(-i (k z-\omega_{0} t)).
\eea
Here $ U $ and $ V$ are the slowly varying complex amplitudes of the forward and backward components, respectively. Also, $\omega$ indicates the carrier frequency. The final nonlinear coupled mode equations (NLCMEs) expressing wave propagation in FBGs with gain and loss are obtained as follows \cite{Pelinovsky2002,Komissarova2019}:
\bes\label{GEq}\bea\label{GEq1}
&&+i\left(\frac{\partial U}{\partial z}+\frac{1}{v}\frac{\partial U}{\partial t}\right)+\left(\kappa\pm g\right)V+\gamma\left(|U|^{2}+2|V|^{2}\right)U\nonumber\\
&&+\xi\left(|V|^{2}+2|U|^{2}\right)V
+\xi\left(U^{2} V^{*}\right)=0,  \\
&&\label{GEq2}-i\left(\frac{\partial V}{\partial z}-\frac{1}{v}\frac{\partial V}{\partial t}\right)+\left(\kappa \mp g\right)U+\gamma\left(|V|^{2}+2|U|^{2}\right)V\nonumber\\
&&+\xi\left(|U|^{2}+2|V|^{2}\right)U
+\xi\left(V^{2} U^{*}\right)=0.
\eea\ees
In Eqs. \eqref{GEq}, the inverse of the coefficient of the second term, $v=c/n_{0}$, represents the group velocity of incident waves, and the coupling parameter ($\kappa$) arises from the real part of the refractive index profiles, denoted as $\kappa=\Delta n_{R} \omega_{0}/c$. Similarly, the contribution from the imaginary part of the refractive index profile is attributed to the gain and loss as $g=\Delta n_{I} \omega_{0}/c$. These parameters play a crucial role in the $\cal PT$-symmetric effecct in the governing equation, given in Eqs. \eqref{GEq}. We wish to note that the coupled mode equations with gain and loss, were previously studied in Ref. \cite{Komissarova2019} addressing the dynamics of optical bistability in different $\cal PT$-symmetric regimes.  Although Porsezian and Senthilnathan \cite{Por2005AIP} have analyzed the emergence of MI in a similar model but without gain and loss, their work considered the conventional version of NLCMEs with a different version of FWM. To explore the impact of the $\cal PT$-symmetric effect in the former system, we consider the $\cal PT$-symmetric NLCMEs with the effect of FWM. The inclusion of the $\cal PT$-symmetric term in Eqs.~\eqref{GEq} reveals a new avenue for the emergence of MI behavior in NLCMEs with the effect of FWM. In this manuscript, we present various new types of behavior for the proposed system, as given in Eqs. \eqref{GEq}, for the first time to the best of our knowledge.

The presence of the $(\pm)$ and $(\mp)$ signs in Eqs. \eqref{GEq} elucidates the direction of light incidence on the given system, respectively, to the left incidence and the right incidence. The nonlinear coefficients for SPM  and FWM are determined by the expressions $\gamma=n_{2} \omega_{0}/c$ and $\xi=\Delta n_{2}\omega_{0}/c$, respectively.
 
\begin{figure*}[t]

  \topinset{(a)}{\includegraphics[scale=0.25]{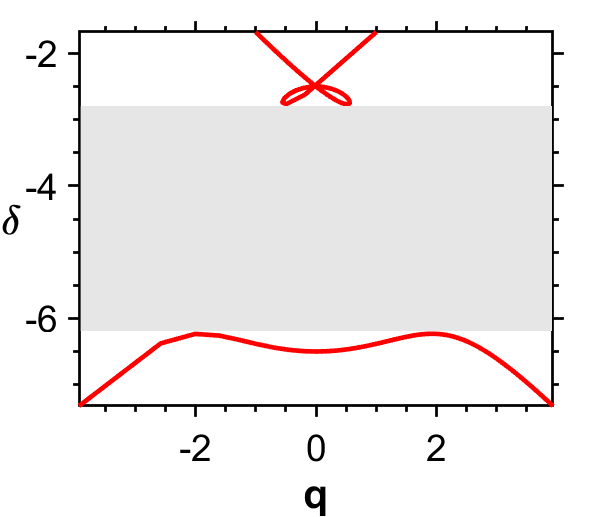}}{0.1in}{-0.3in}
  % {$g=0$};
  %{$\xi=-1$};
  \topinset{(b)}{\includegraphics[scale=0.25]{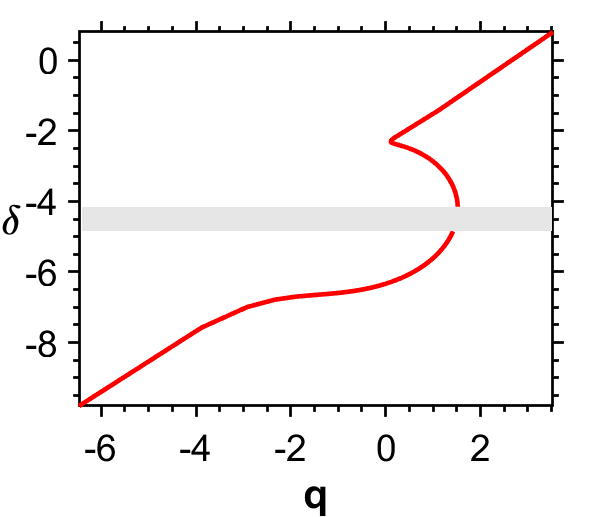}}{0.1in}{-0.3in}
  %{$g=0.5$};
  \topinset{(c)}{\includegraphics[scale=0.25]{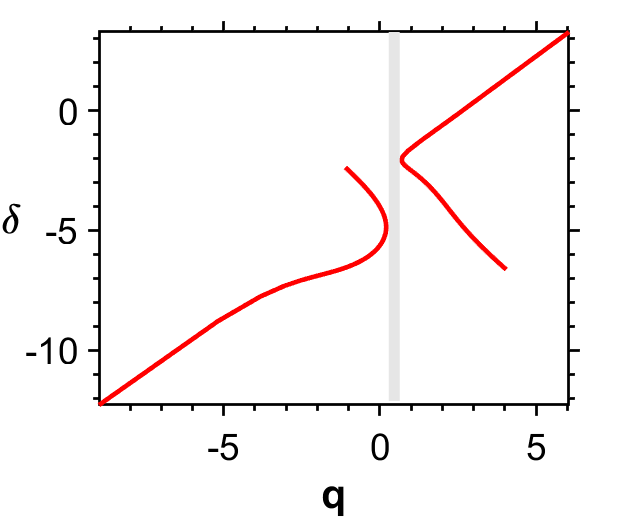}}{0.1in}{-0.3in}
  \topinset{(d)}{\includegraphics[scale=0.25]{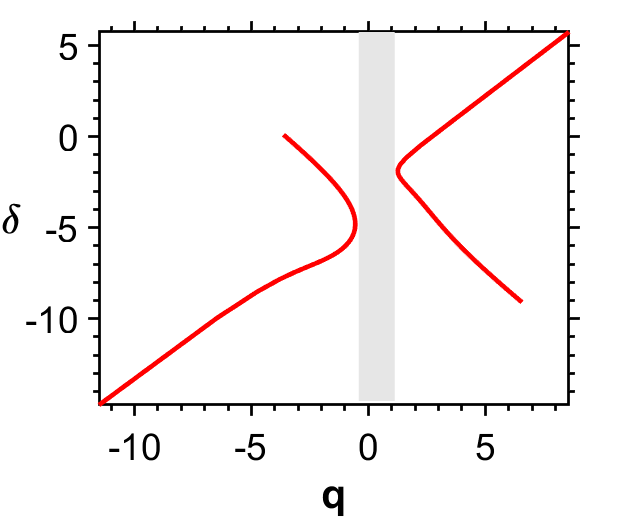}}{0.1in}{-0.3in}\\
   \topinset{(e)}{\includegraphics[scale=0.24]{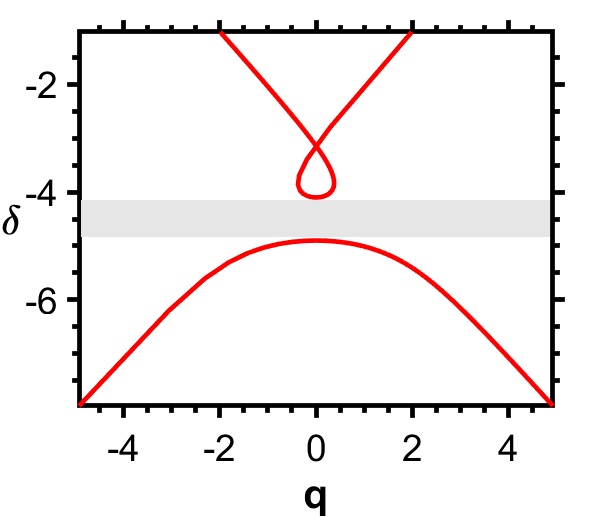}}{0.1in}{-0.3in}
   % anchor=center,yshift=-0.7cm,font=\color{red}] {$\xi=-0.2$};
  \topinset{(f)}{\includegraphics[scale=0.25]{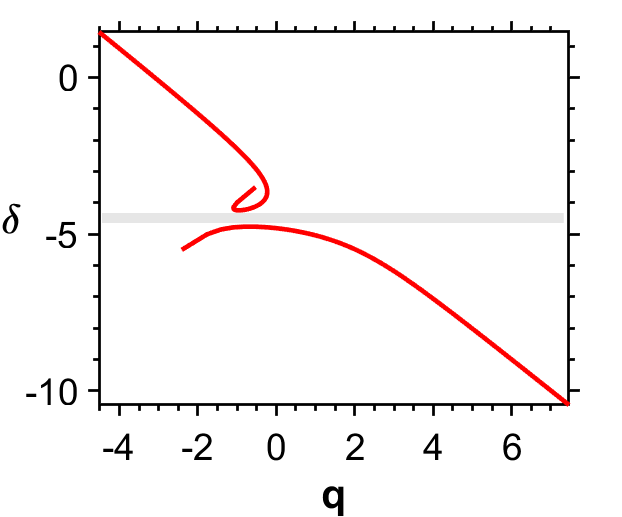}}{0.1in}{-0.3in}
 \topinset{(g)}{\includegraphics[scale=0.25]{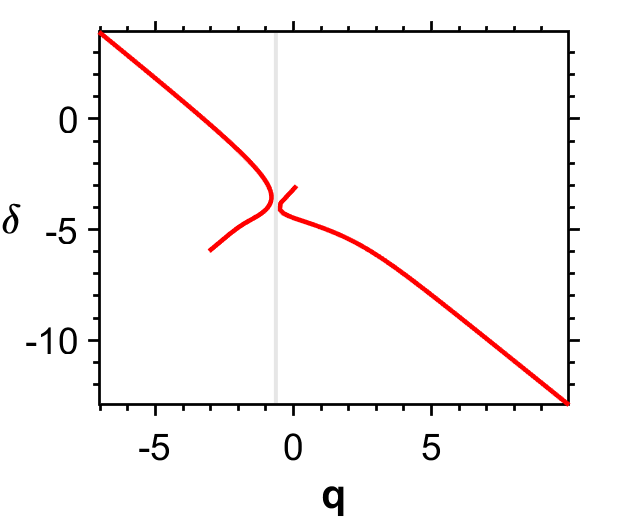}}{0.1in}{-0.3in}
 \topinset{(h)}{\includegraphics[scale=0.25]{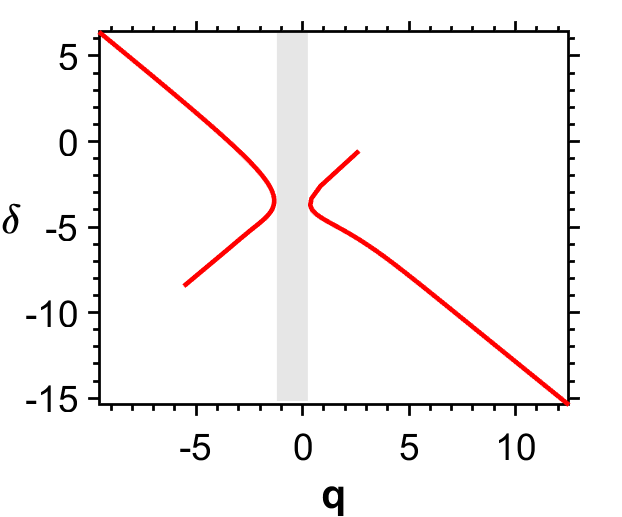}}{0.1in}{-0.3in}\\
  \topinset{(i)}{\includegraphics[scale=0.24]{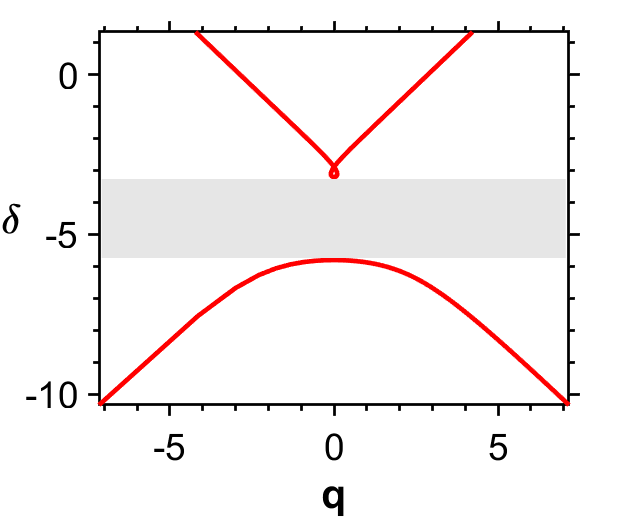}}{0.1in}{-0.3in}
  %\node[left=of img6, node distance=0cm, rotate=90, anchor=center,yshift=-0.7cm,font=\color{red}] {$\xi=0.1$};
 \topinset{(j)}{\includegraphics[scale=0.25]{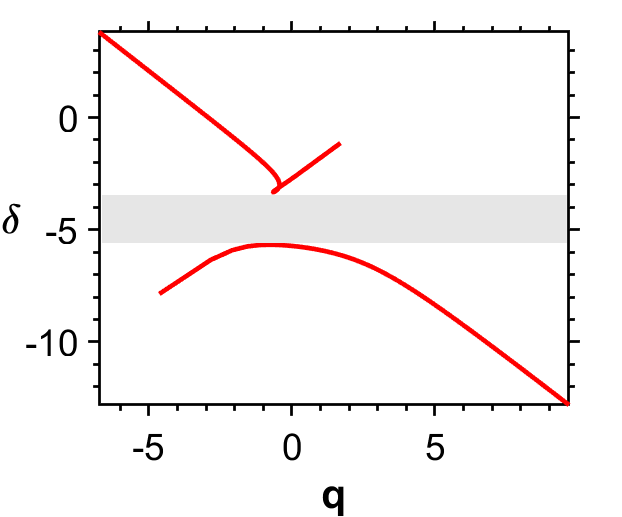}}{0.1in}{-0.3in}
\topinset{(k)}{\includegraphics[scale=0.25]{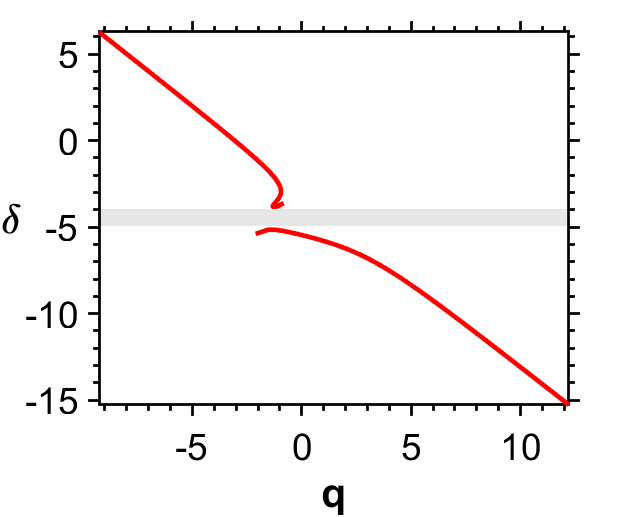}}{0.1in}{-0.3in}
\topinset{(l)}{\includegraphics[scale=0.25]{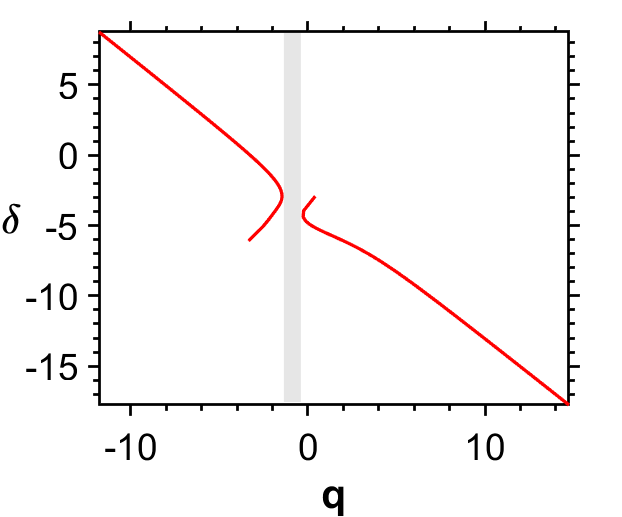}}{0.1in}{-0.3in}\\
 \topinset{(m)}{\includegraphics[scale=0.24]{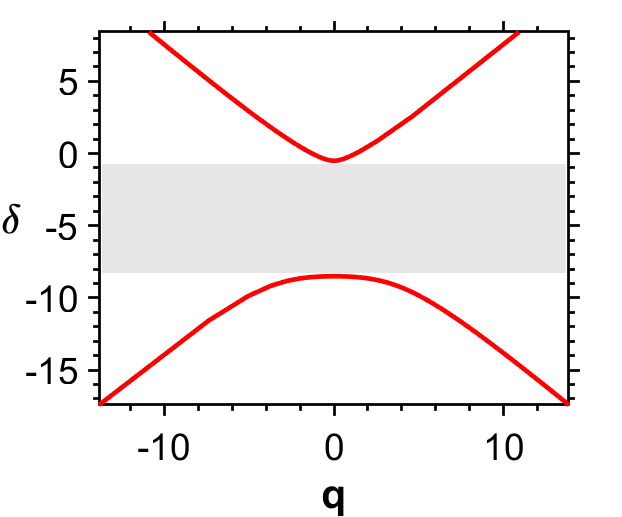}}{0.1in}{-0.3in}
\topinset{(n)}{\includegraphics[scale=0.25]{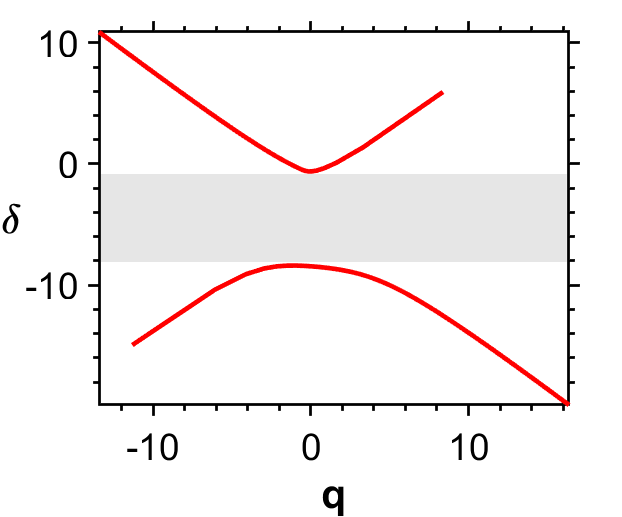}}{0.1in}{-0.3in}
\topinset{(0)}{\includegraphics[scale=0.25]{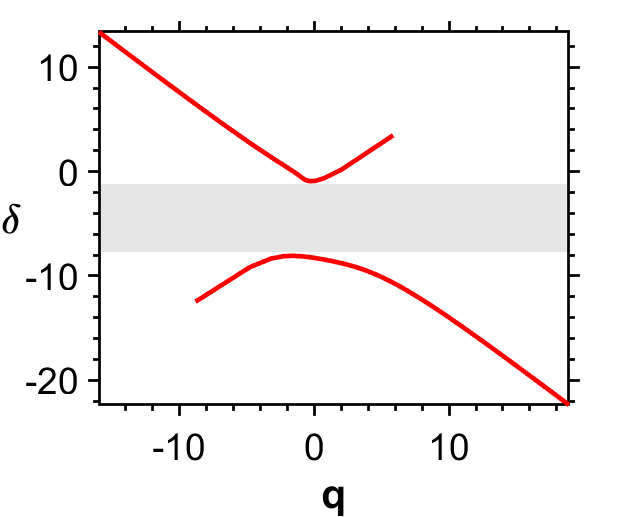}}{0.1in}{-0.3in}
\topinset{(p)}{\includegraphics[scale=0.25]{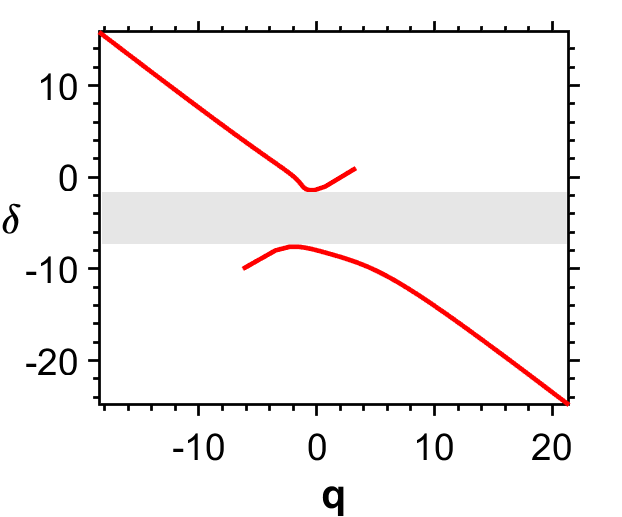}}{0.1in}{-0.3in}
    	\caption{(Color online) Nonlinear dispersion relation of $\cal PT$-symmetric FBG system with the effect of FWM is depicted between $\delta$ and $q$ for the conventional case (first column) with $\xi=-1$ (a - d), unbroken $\cal PT$-symmetric regime (second column) with $\xi=-0.2$ (e - h), at the $\cal PT$-symmetric threshold (third column) with $\xi= 0.1$ (i - l), and broken $\cal PT$-symmetric regime (fourth column) with $\xi= 1$ (m -
    	 p). The values of the system parameters are $P=1.5$, $\kappa=1$, and $\gamma=2$. The incidence of light is left in this case.}\label{Figure1}
\end{figure*}

\section{The characteristics of dispersion relation} \label{MI}

As a first step, we aim to investigate the dispersion relation of the  system \eqref{GEq} in all possible cases, encompassing both the conventional and various $\cal {PT}$-symmetric regimes. For this purpose, we consider counter-propagating CW solutions as follows:
\bes\label{CW}\bea
U= a \exp(i (q z - \delta v t)), \\
V= b \exp(i (q z - \delta v t)).
\eea\ees
In the solutions given above, Eqs. \eqref{CW}, the parameters $a$ and $b$ denote the real constant amplitudes of the forward and backward waves, respectively and the detuning frequency is denoted by the term $\delta$. The total power of the grating structure is defined as $P = a^{2} + b^{2}$. We introduce a term representing the ratio between these amplitudes as $f = b/a$, so that
\bea 
a = \sqrt{\frac{P}{(1+f^2)}}, \qquad \text{and} \qquad b = \sqrt{\frac{Pf^2}{(1+f^2)}}.
\eea
Upon substituting Eqs.\eqref{CW} into the system \eqref{GEq}, one obtains the following mathematical expression for the nonlinear dispersion relations:
\bes \label{deltaq}\bea
q=-\frac{(\kappa+\xi P)}{2f}\left(1-f^{2}\right)\pm\cfrac{g}{2f}\left(1+f^{2}\right)-\cfrac{\gamma~P~\epsilon_{1}}{2}, \qquad\\
\delta=-\cfrac{\kappa}{2f}\left(1+f^{2}\right)\pm\cfrac{g}{2f}\left(1-f^{2}\right)-\cfrac{3\gamma P~}{2}-\cfrac{\xi P~\epsilon_{2}}{2f}. \qquad\label{Eq:dr2}
\eea \ees
Here, the notations $ \epsilon_{1,2}$ are used to describe the fractional terms $\frac{(1-f^{2})}{(1+f^{2})}$ and $\frac{(6f^{2}+f^{4}+1)}{(1+f^{2})}$, respectively. The above expressions \eqref{deltaq} help us understand the dispersion characteristics of the proposed model \eqref{GEq} by tuning the system parameters, such as SPM and FWM nonlinearities, as well as gain and loss.

It is to be noted that in the entire manuscript, the value of coupling coefficient is scaled to be one (i.e. $\kappa=1$) and based on this value, the four important regimes are classified as the conventional case ($g=0$), unbroken $\cal{PT}$ -symmetric ($g=0.5$), at the threshold ($g=1$), and broken $\cal{PT}$ -symmetric ($g=1.5$) regimes unless otherwise specified.

In Eq. \eqref{deltaq}, the parameter $f$ is crucial because when it is negative $(f<0)$, it corresponds to anomalous dispersion, while when $f$ is positive $(f>0)$, it represents normal dispersion regime. The specific values of $f$, namely $f=\pm 1$, correspond to the two edges of the photonic band gap in the linear case, which is not reported here. Specifically, $f=-1$ reflects the top of the photonic band gap, while $f=+1$ represents the bottom of the photonic band gap. From the above dispersion relations, given in Eq. \eqref{deltaq}, our next goal is to examine how the characteristics of the dispersion relations dynamically change under the influence of FWM across the four different $\mathcal{PT}$-symmetric conditions. To achieve this, we assign four different values to the FWM parameter, ranging from $\xi=-1$ to $1$ in the case of left light incidence.

We now consider the case of FWM with a very low value, like $\xi=-1$, under different $\mathcal{PT}$-symmetric regimes. In the conventional case with $g=0$, there is an unusual propagation of forward and backward wave vectors, as depicted in Fig.~\ref{Figure1}(a). To the best of our knowledge, this type of unusual dispersion curve is reported for the first time, in the literature. Specifically, the forward wave vector exhibits a double-loop structure, while the backward wave vector displays a double-hump structure. This novel dispersion curve occurs in a considerably broad bandgap, anticipating the formation of gap solitons, in the considered grating system.

Upon further increasing the value of $g$ ($g=0.5$), i.e., below the $\mathcal{PT}$-symmetric threshold, the novel double-loop, and double-hump structures undergo deformation and they attempt to align in a single direction, facing each other with a reduced band gap, as illustrated in Fig.~\ref{Figure1}(b). These scenarios undergo a complete transformation when the system reaches the exceptional point at $g=1$ (also known as the $\mathcal{PT}$ -symmetric threshold). Interestingly, the dispersion curve rotates 90 degrees clockwise, as depicted in Fig. \ref{Figure1}(c). Furthermore, in this case, it is evident that the bandgap gets decreased. In the broken $\mathcal{PT}$-symmetric regime, the band gap gets increased as depicted in Fig. \ref{Figure1}(d), when compared to the previous case.

Next, we consider the case of $\xi=-0.2$ under different  $\mathcal{PT}$-symmetric conditions. In the conventional case, as illustrated in Fig.~\ref{Figure1}(e), we observe the typical dispersion curve exhibiting a single loop structure on the upper dispersion curve with a reduced bandgap. The dispersion curve shown in Fig. \ref{Figure1}(e) gets slightly deformed when the system operates in the unbroken $\cal PT$-symmetric regime, as illustrated in Fig.~\ref{Figure1}(f). Also, the cases of $\cal PT$-symmetric threshold and broken $\cal PT$-symmetric regimes (Figs. \ref{Figure1}(g) and (h)) reveal the same dynamics in the dispersion curves as observed in Figs. \ref{Figure1} (g) and (h) when the value of the FWM is $\xi=-1$ with a difference being that the dispersion curves get shifted. 

In the case of FWM with a positive low value like $\xi=0.1$ the dispersion curves mimic the same dynamics as in all the cases of $\xi=-0.2$ with increased bandgap except for the $\cal{PT}$-symmetric threshold which exhibits a rotation of the dispersion curves, see  Figs.~\ref{Figure1}(i) to  \ref{Figure1}(l). On the other hand, for $\xi=1$, in sharp contrast to the former three cases, the dispersion curves do not exhibit any unusual characteristics  such as a loop formation or the rotation of the curves [see Figs. \ref{Figure1}(m) to \ref{Figure1}(p).
 
When comparing the above dispersion results for $g=0$ with the dispersion relation of the conventional NLCMEs with FWM  reported in Ref. \cite{Por2005AIP}, several distinct features in the conventional case with $g=0$ become apparent. Firstly, for $\xi=1$, there is an unusual forward wave vector exhibiting a double-loop structure, while the backward wave vectors display a two-hump structure. Secondly, our dispersion relation does not invert for negative values of $\xi$. Additionally, the non-identical band structure of forward and backward wave vectors remains distinct even for $\xi=1$. 

\begin{figure*}[t]
 \topinset{\color{white}(a)}{\includegraphics[scale=0.28]{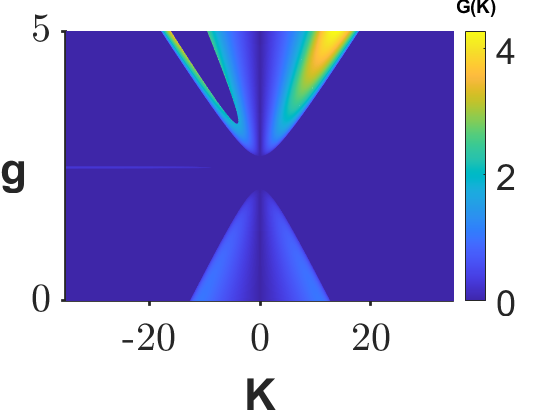}}{0.1in}{-0.3in}
  % \node[left=of img8, node distance=0cm, rotate=90, anchor=center,yshift=-0.7cm,font=\color{black}]{Left Incidence}; 
 \topinset{\color{white}(b)}{\includegraphics[scale=0.28]{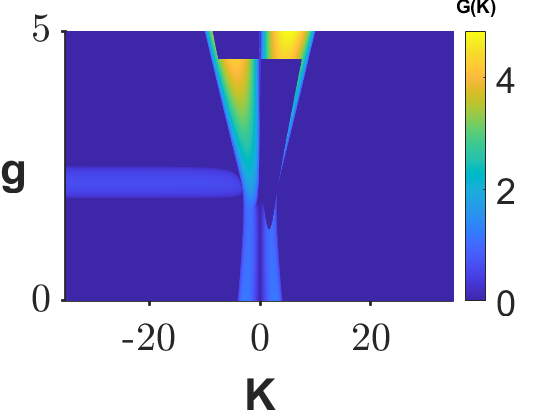}}{0.1in}{-0.3in}
 \topinset{\color{white}(c)}{\includegraphics[scale=0.28]{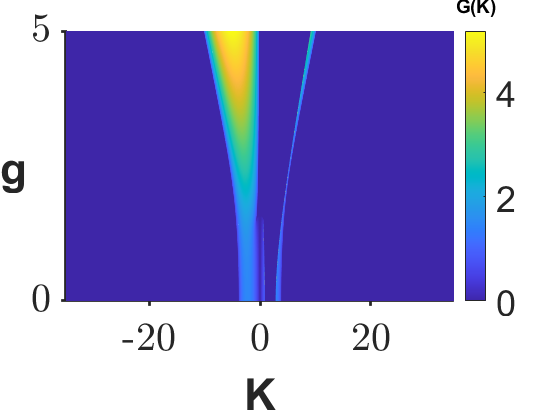}}{0.1in}{-0.3in}
 \topinset{\color{white}(d)}{\includegraphics[scale=0.28]{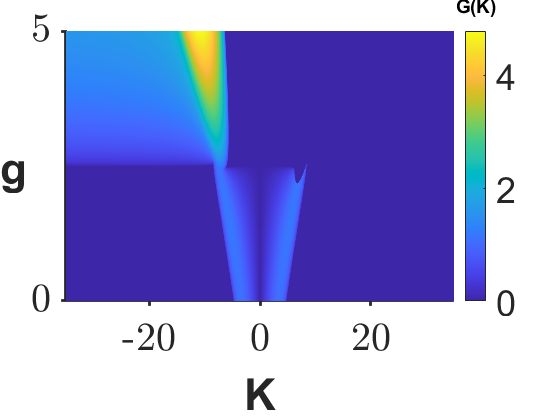}}{0.1in}{-0.3in}\\

 \topinset{\color{white}(e)}{\includegraphics[scale=0.27]{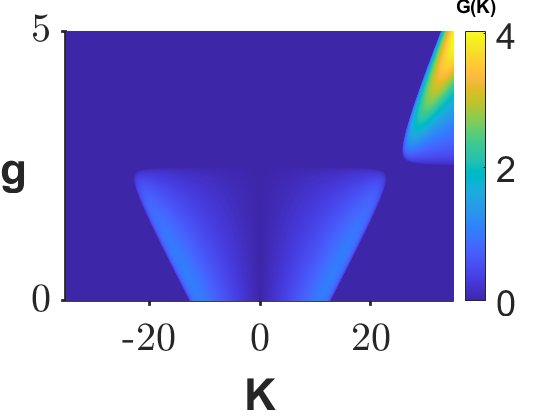}}{0.1in}{-0.3in}
   %  \node[left=of img9, node distance=0cm, rotate=90, anchor=center,yshift=-0.7cm,font=\color{black}] {Right Incidence};
\topinset{\color{white}(f)}{\includegraphics[scale=0.28]{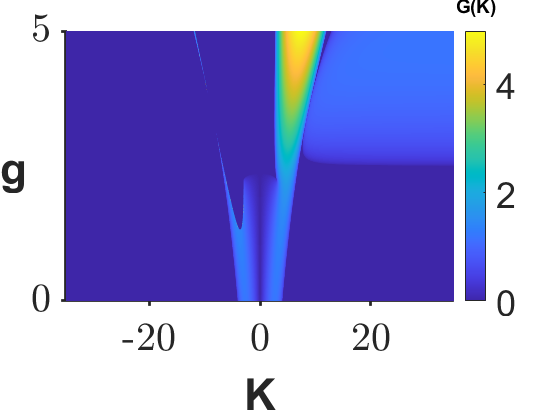}}{0.1in}{-0.3in}
\topinset{\color{white}(g)}{\includegraphics[scale=0.28]{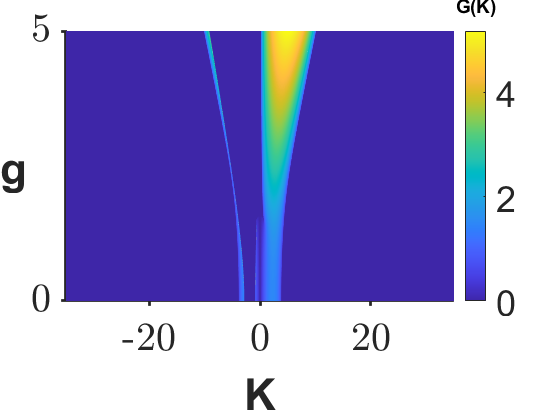}}{0.1in}{-0.3in}
\topinset{\color{white}(h)}{\includegraphics[scale=0.28]{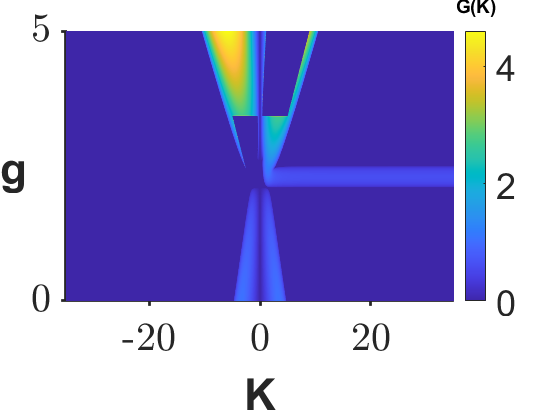}}{0.1in}{-0.3in}
\caption{(Color online) MI gain spectrum in the anomalous dispersion regime with FWM $\xi=1$ for left incidence (top panels) and right incidence (bottom panels) for (a) and (e) $f=-0.1$, (b) and (f) $f=-0.5$, (c) and (g) $f=-1$, and (d) and (h) $f=-3$. The other parameters are $P=1.5$, $\kappa=1$, and $\gamma=2$.}\label{Figure2}
\end{figure*}
\section{The dynamics of modulational instability}
\label{Sec:III}
To study the MI dynamics, we employ the standard analytical procedure, namely the linear stability analysis (LSA). To apply this LSA method, infinitesimal perturbations are imposed on the continuous wave (CW) state, resulting in the exponential growth of its perturbed amplitude. Following the general procedure, let us introduce the forward and backward wave components with perturbations as
\bes\label{CW1}\bea 
U = (a+ \Psi_{1}(z,t)) \exp(i(qz-\delta v t)),\\
V = (b+ \Psi_{2}(z,t)) \exp(i(qz-\delta v t)).
\eea\ees
Note that the perturbed amplitudes of the forward and backward waves are considered in such a way that $|\Psi_{1,2}|$ $\ll a, b$. Upon substitution of Eqs. \eqref{CW1} into Eqs. \eqref{GEq} and linearizing the perturbations with respect to $\Psi_{1,2}$, one can arrive at the following  continuous wave (CW) equations for the forward and backward wave components:
\begin{widetext}
\bes\label{LEQ}\bea
&&i \Psi_{1,z}+\left(\frac{i}{v}\right)\Psi_{1,t}-\eta \left(f\Psi_{1}-\Psi_{2}\right)+ \Upsilon_{0} \left[ \gamma \left(\Psi_{1}+\Psi_{1}^{*}+2f(\Psi_{2}+\Psi_{2}^{*})\right)+\xi \left((f-f^{3}) \Psi_{1}+2 f \Psi_{1}^{*}\right) \right] +\xi P (2 \Psi_{2}+\Psi_{2}^{*})=0,\label{LEQ1}\qquad\\
&& -i\Psi_{2,z}+\left(\frac{i}{v}\right)\Psi_{2,t}+\eta_{1}(\Psi_{1}-\frac{1}{f}\Psi_{2})+\Upsilon_{0} \left[\gamma(2 f \left(\Psi_{1}+\Psi_{1}^{*})+f^{2}(\Psi_{2}+\Psi_{2}^{*})\right)+\xi \left((f-\Upsilon_{1} )\Psi_{2}+2 f \Psi_{2}^{*}\right) \right] +2\xi  \Psi_{1}+\Psi_{1}^{*})=0.\label{LEQ2}\qquad
\eea\ees
\end{widetext}
Here the terms $\Upsilon_{0,1}$ are expressed as $P/(1+f^{2})$ and $1/f$, respectively. Additionally, $\eta=\kappa\pm g$, and $\eta_{1}=\kappa\mp g$. Then, we introduce the Fourier components of the perturbed forward and backward continuous wave (CW) amplitudes $\Psi_{1,2}(z,t)$ as:
\bes\label{PertEq}\bea\label{PertEq1}
&&\Psi_{1}(z,t)=\mbox{u}_{+} e^{i(K z-\Omega v t)}+\mbox{u}_{-} e^{-i(K z-\Omega v t)},\\
&&\Psi_{2}(z,t)=\mbox{v}_{+} e^{i(K z-\Omega v t)}+\mbox{v}_{-} e^{-i(K z-\Omega v t)}.\label{PertEq2}
\eea\ees
In Eqs. \eqref{PertEq}, $\mbox{u}_{+}$ and $\mbox{v}_{+}$ represent forward propagation, while $\mbox{u}_{-}$ and $\mbox{v}_{-}$ represent backward propagation. Also, $K$ and $\Omega$ denote the wavenumber and frequency of the perturbation, respectively.
 Substituting the solutions \eqref{PertEq1} and \eqref{PertEq2} into the governing equations \eqref{LEQ} the resulting equations with respect to the perturbed amplitudes $\mbox{u}_{+}$ and $\mbox{u}_{-}$ and $\mbox{v}_{+}$ and $\mbox{v}_{-}$ yield four homogeneous equations in the following matrix form:
\bea\label{matrix}
[A]\times[y]^{T}=0, \qquad\qquad y^{T}=(\mbox{u}_{+},\mbox{v}_{+},\mbox{u}_{-},\mbox{v}_{-})\
\eea
where, $A$ is a $4\times4$ matrix with the following elements:
\bea
&&a_{11}= -f (\kappa\pm g)-K +\Upsilon_{0}(\gamma  +f \xi -f^{3} \xi)+\Omega,\nonumber\\
&&a_{12}=\Upsilon_{0}\left(\gamma+2f \xi\right),\nonumber\\ 
&&a_{13}=\kappa\pm g+2 f \Upsilon_{0}\gamma+2 P \xi,
\nonumber\\
&&a_{14}=2 f \Upsilon_{0}\gamma+P \xi,\nonumber\\
&&a_{21}=\Upsilon_{0}\gamma+2 f \Upsilon_{0}\xi,\nonumber\\
&&a_{22}=-f(\kappa\pm g)+K +\Upsilon_{0}\gamma+\xi(f-f^{3})\Upsilon_{0}-\Omega,\nonumber\\
&&a_{23}=2f\Upsilon_{0}\gamma+P\xi+\nonumber\\
&&a_{24}=\kappa\pm g+2 f \Upsilon_{0}\gamma+2 P\xi,\nonumber\\
&& a_{31}=\kappa\mp g +2 f \Upsilon_{0}\gamma+2 P \xi,\nonumber\\
&&a_{32}=2 f\Upsilon_{0}\xi+P\xi,\nonumber\\
&&a_{33}=-\Upsilon_{1}(\kappa\mp g)+K + \Upsilon_{0}(f^{2} \gamma-\xi+ f \xi)+\Omega \nonumber\\
&&a_{34}=\Upsilon_{0}(f^{2} \gamma+2f\xi),\nonumber\\
&&a_{41}=\Upsilon_{0}(2f\gamma)+P\xi,\nonumber\\
&&a_{42}=\kappa\mp g+2 f \Upsilon_{0}\gamma+2 P\xi,
\nonumber\\
&&a_{43}=\Upsilon_{0}(f^{2} \gamma+2 f \xi),\nonumber\\
&&a_{44}=-\Upsilon_{1}(\kappa\mp g)-K + \Upsilon_{0}(\gamma  f^{2}-\xi+f\xi)-\Omega.
\eea
It is important to emphasize that the  matrix $A$ yields non-trivial solutions when its determinant reaches zero, resulting in a quartic polynomial equation for $\Omega$. This quartic equation produces four different solutions,
for which, the growth rate of the MI , denoted as $G(K)$, can be computed using the relationship $G(K) = \text{Im}(\Omega_{\text{max}})$, where $\Omega_{\text{max}}$ represents the largest imaginary part among the four branches.
\begin{figure}[t]
   \topinset{(a)}{\includegraphics[scale=0.22]{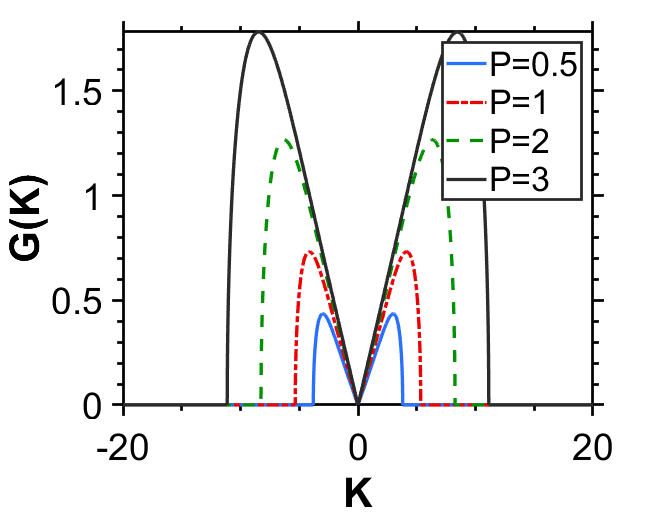}}{0.1in}{-0.3in}
  \topinset{(b)}{\includegraphics[scale=0.21]{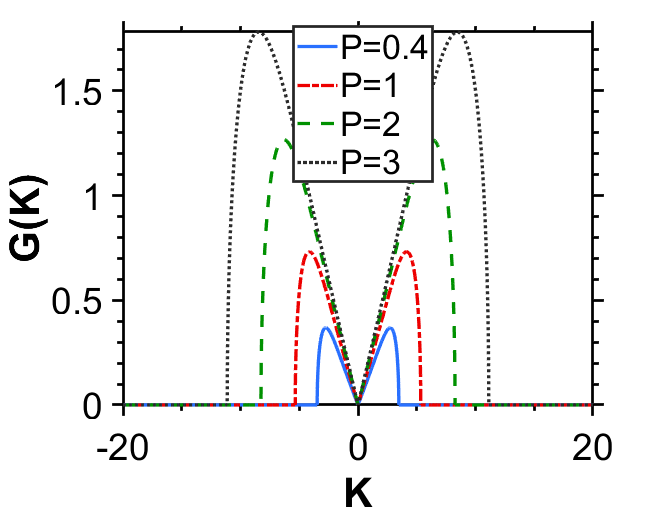}}{0.1in}{-0.3in}\\
  \topinset{(c)}{\includegraphics[scale=0.22]{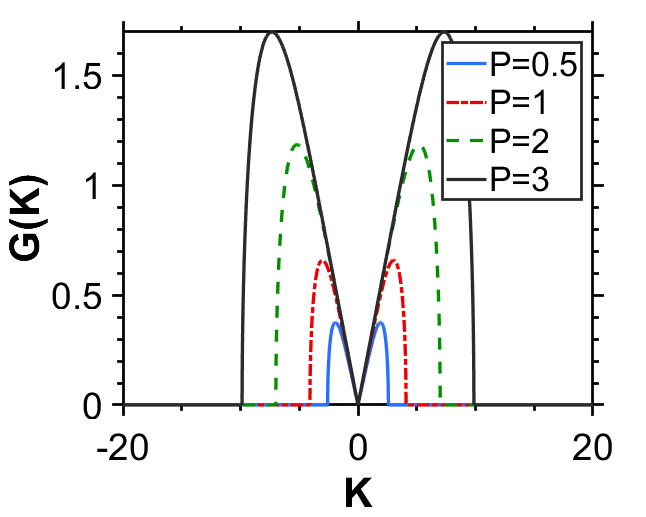}}{0.1in}{-0.3in}
\topinset{(d)}{\includegraphics[scale=0.22]{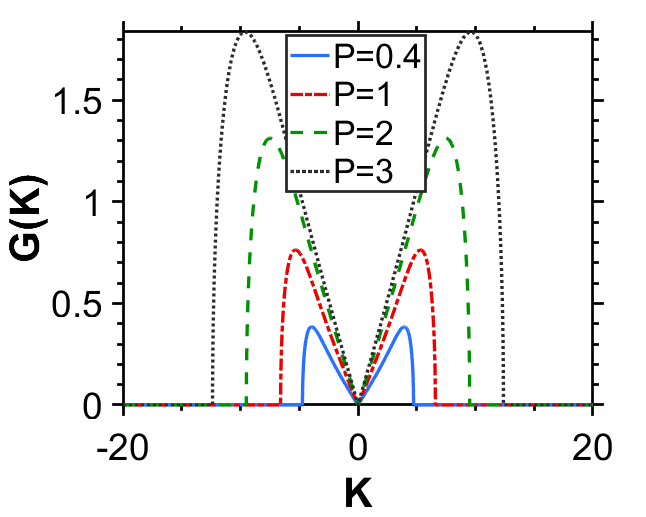}}{0.1in}{0.3in}\\
 \topinset{(e)}{\includegraphics[scale=0.22]{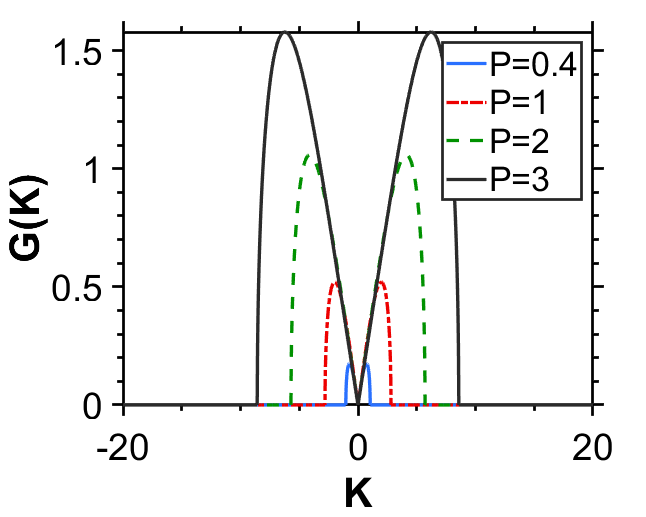}}{0.1in}{-0.3in}
\topinset{(f)}{\includegraphics[scale=0.22]{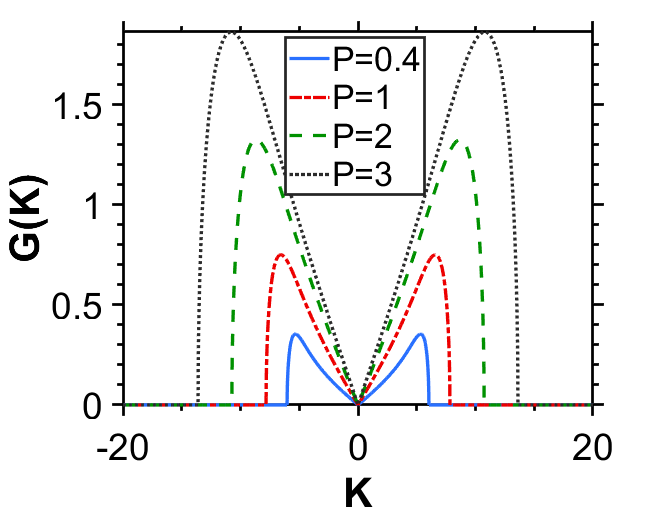}}{0.1in}{0.3in}\\
 \topinset{(g)}{\includegraphics[scale=0.22]{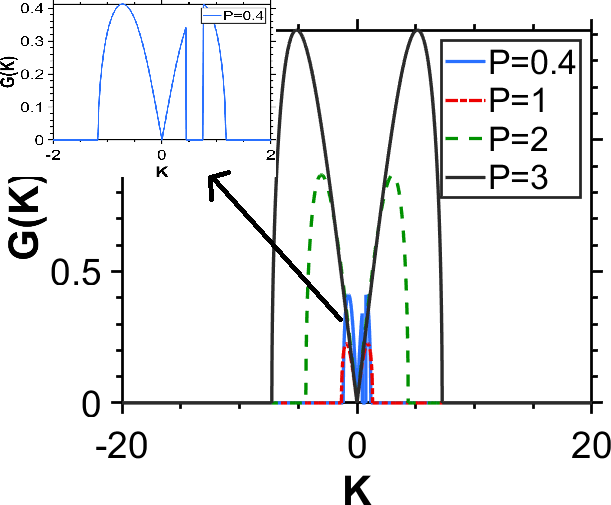}}{0.1in}{0.1in}
\topinset{(h)}{\includegraphics[scale=0.22]{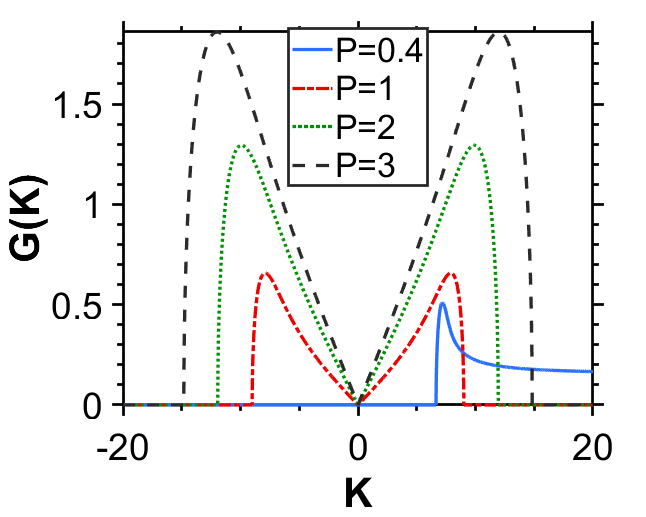}}{0.1in}{0.3in}
\caption{(Color online) Plots show the role of power on the MI gain spectrum in the anomalous dispersion regime for left incidence (left panels) and right incidence (right panels) for different $\cal PT$-symmetric regimes, including (a-b) conventional, (c-d) unbroken $\cal PT$-symmetric, (e-f) exceptional point, and (g-h) broken $\cal PT$-symmetric regime. The values of other parameters are $f=-0.2$, $\kappa=1$, $\gamma=2$, and $\xi=1$.}\label{Figure3}
\end{figure}

\begin{figure}
\topinset{(a)}{\includegraphics[scale=0.21]{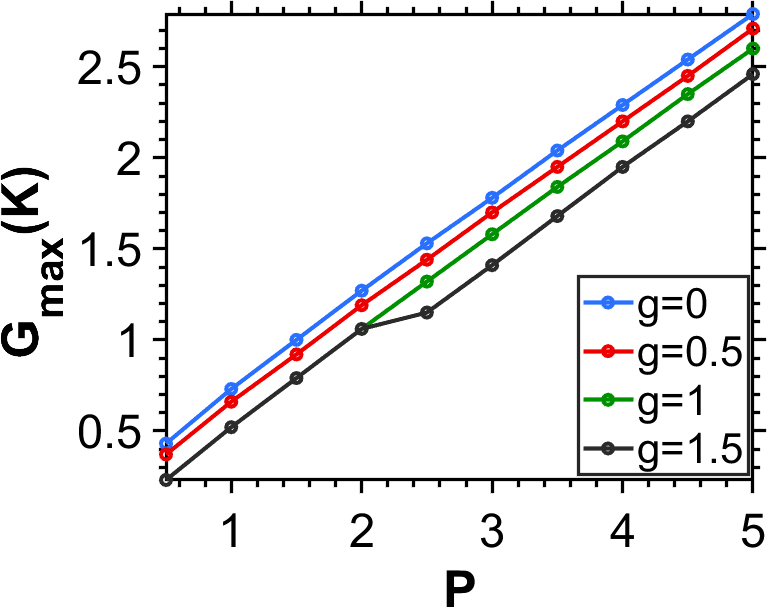}}{0.1in}{-0.3in}
\topinset{(b)}{\includegraphics[scale=0.21]{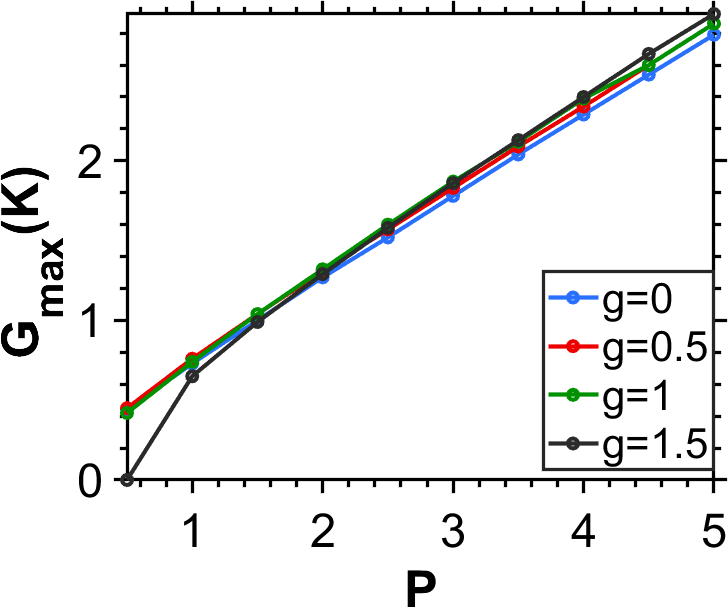}}{0.1in}{-0.3in}
\caption{(Color online) The maximum gain of the MI spectra versus $P$ in the anomalous dispersion regime in the positive wavenumber region is shown  for (a) left incidence and (b) right incidence for four different $\cal PT$-symmetric conditions. The parameters are the same as given in Fig. \ref{Figure3}.}
\end{figure}

\begin{figure}[t]
 \topinset{(a)}{\includegraphics[scale=0.22]{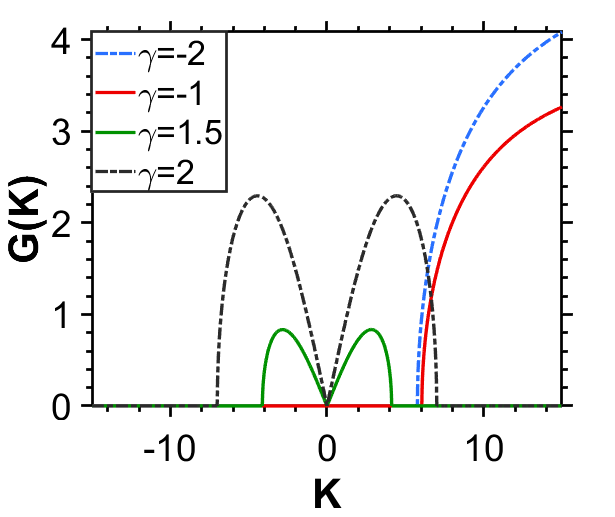}}{0.1in}{0.3in}
    \topinset{(b)}{\includegraphics[scale=0.21]{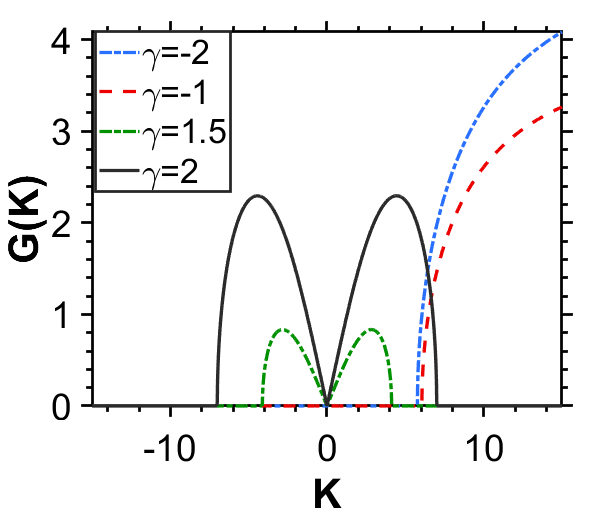}}{0.1in}{0.3in}\\
  \topinset{(c)}{\includegraphics[scale=0.22]{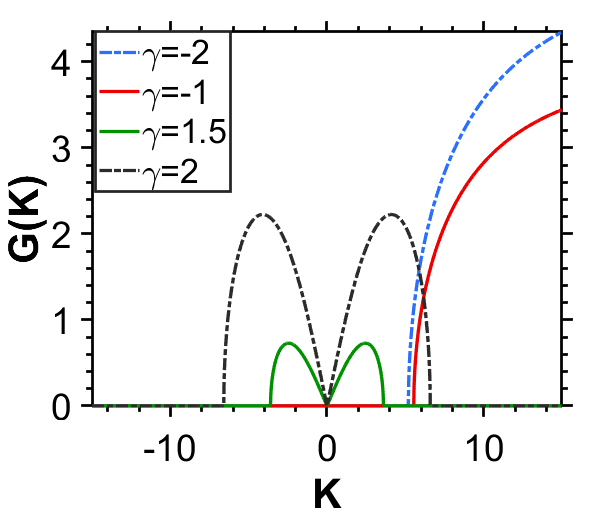}}{0.1in}{0.3in}
\topinset{(d)}{\includegraphics[scale=0.22]{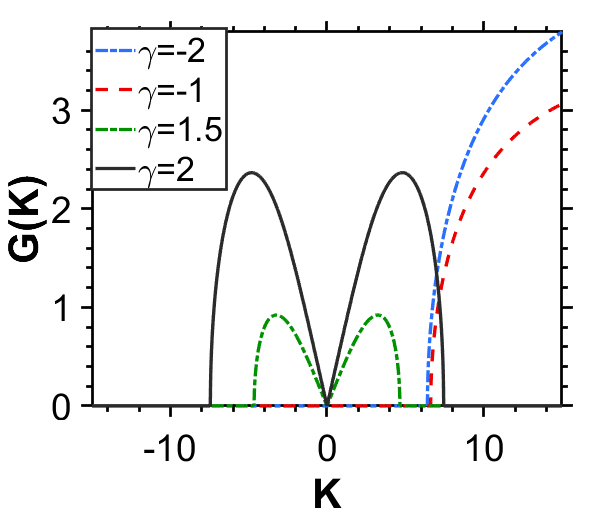}}{0.1in}{0.3in}\\
 \topinset{(e)}{\includegraphics[scale=0.22]{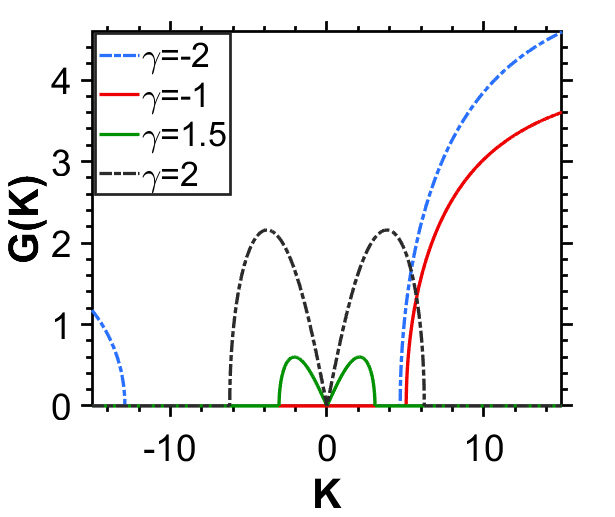}}{0.1in}{0.3in}
\topinset{(f)}{\includegraphics[scale=0.22]{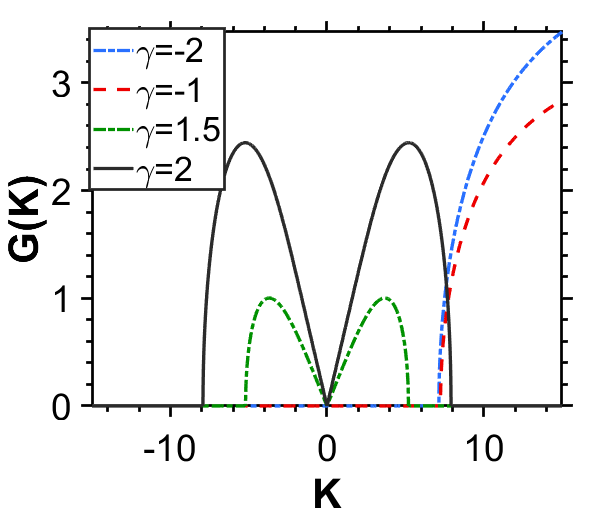}}{0.1in}{0.3in}\\
 \topinset{(g)}{\includegraphics[scale=0.22]{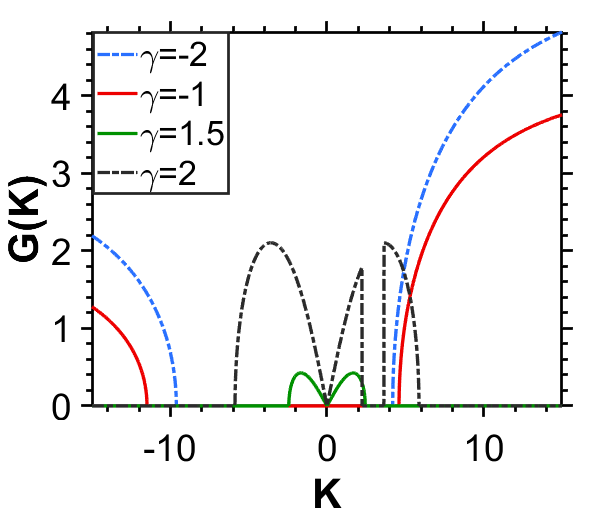}}{0.1in}{0.3in}
\topinset{(h)}{\includegraphics[scale=0.22]{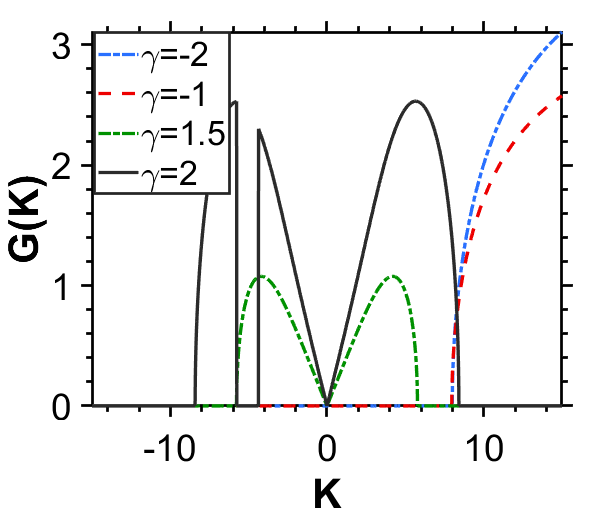}}{0.1in}{0.3in}
\caption{(Color online) Plots depict how SPM influences the MI gain spectrum in the anomalous dispersion regime for left incidence (left panels) and right incidence (right panels), and for (a-b) conventional, (c-d) unbroken $\cal PT$-symmetric, (e-f) exceptional point, and (g-h) broken $\cal PT$-symmetric regime with $f=-0.5$, $\kappa=1$, $P=3$, and $\xi=1$. }\label{Figure4}
\end{figure}
\begin{figure*}
 \topinset{\color{white}(a)}{\includegraphics[scale=0.28]{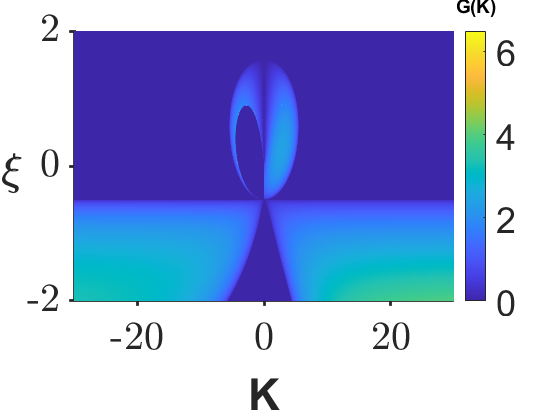}}{0.1in}{-0.3in}
\topinset{\color{white}(b)}{\includegraphics[scale=0.28]{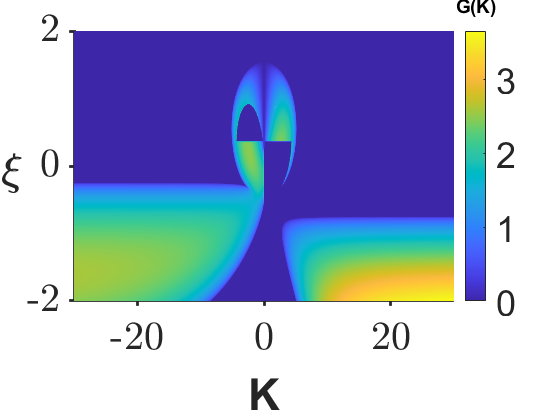}}{0.1in}{-0.3in}
\topinset{\color{white}(c)}{\includegraphics[scale=0.28]{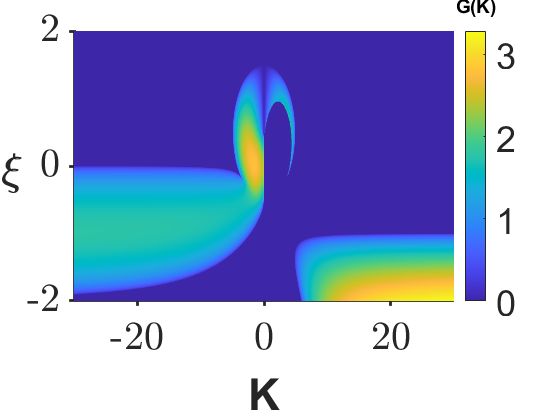}}{0.1in}{-0.3in}
\topinset{\color{white}(d)}{\includegraphics[scale=0.28]{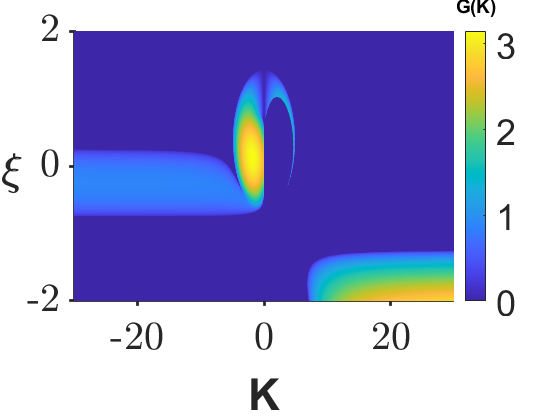}}{0.1in}{-0.3in}\\
 \topinset{\color{white}(e)}{\includegraphics[scale=0.28]{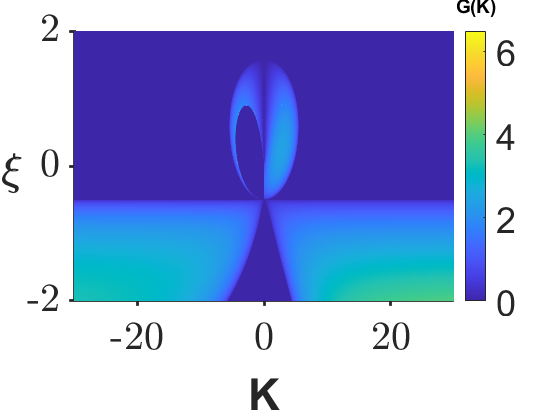}}{0.1in}{-0.3in}
\topinset{\color{white}(f)}{\includegraphics[scale=0.28]{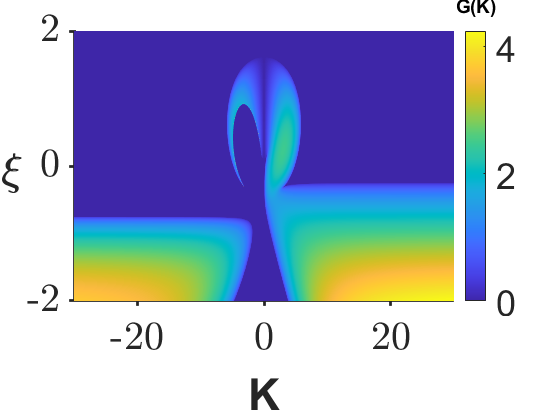}}{0.1in}{-0.3in}
\topinset{\color{white}(g)}{\includegraphics[scale=0.28]{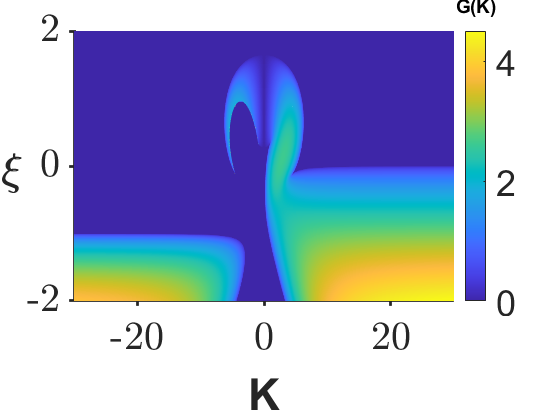}}{0.1in}{-0.3in}
\topinset{\color{white}(h)}{\includegraphics[scale=0.28]{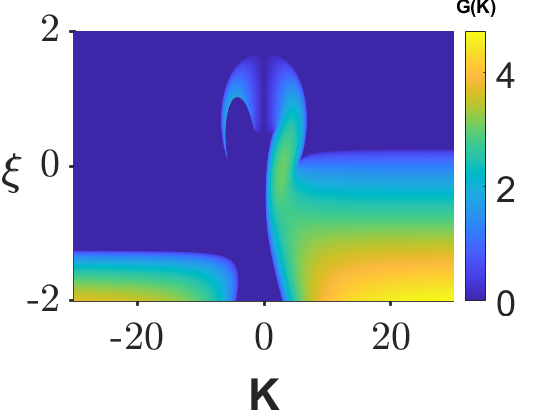}}{0.1in}{-0.3in}
\caption{(Color online) MI is plotted as a function of FWM in both the left (top panels) and right (bottom panels) incidences of the proposed system for (a) and (e) conventional, (b) and (f) unbroken $\cal PT$-symmetric, (c) and (g) exceptional point, and (d) and (h) broken $\cal PT$-symmetric regimes with $f=-0.5$, $P=2$, $\kappa=1$, and $\gamma=2$. }\label{Figure5}
\end{figure*}
\begin{figure}[ht]
 \topinset{(a)}{\includegraphics[scale=0.21]{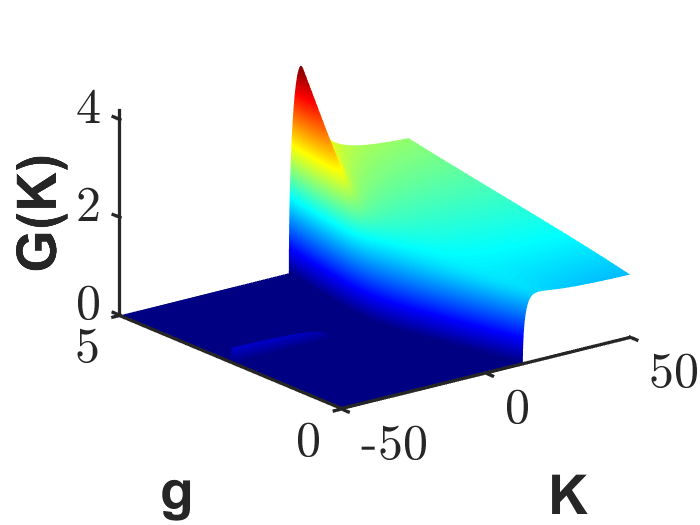}}{0.1in}{-0.3in}
 \topinset{(b)}{\includegraphics[scale=0.21]{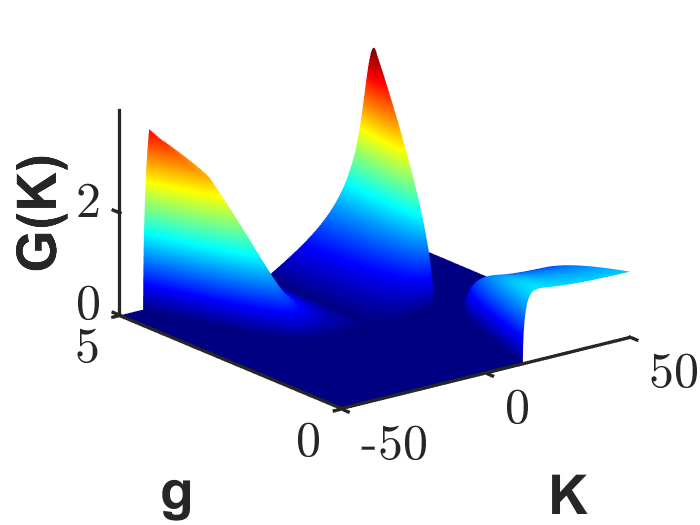}}{0.1in}{-0.3in}\\
\topinset{(c)}{\includegraphics[scale=0.21]{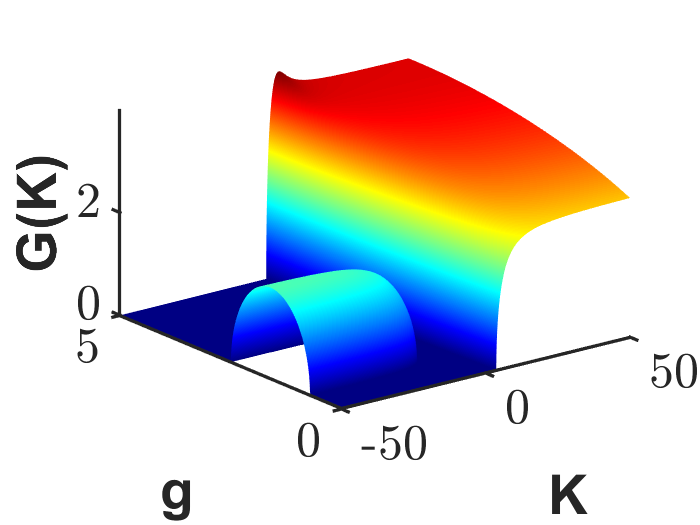}}{0.1in}{-0.3in}
\topinset{(d)}{\includegraphics[scale=0.21]{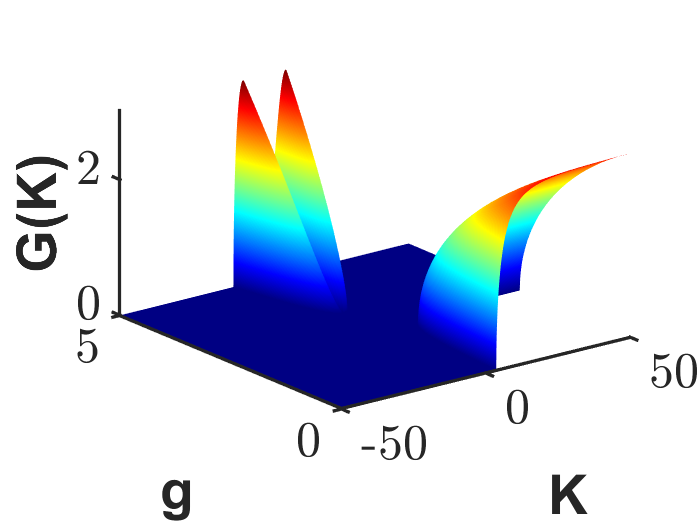}}{0.1in}{-0.3in}\\
\topinset{(e)}{\includegraphics[scale=0.21]{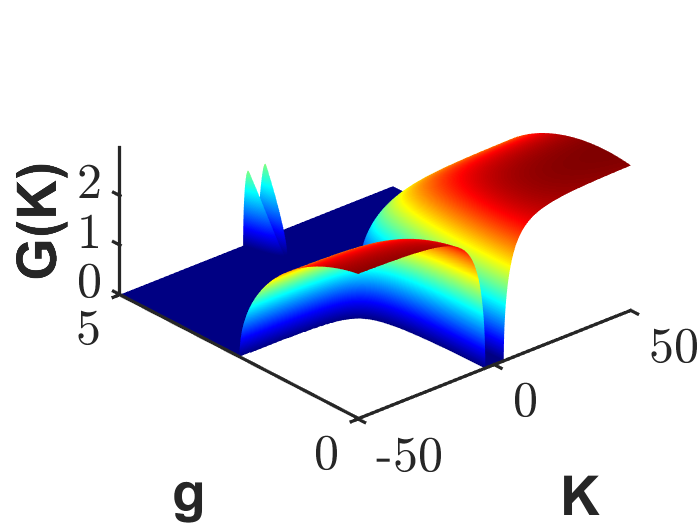}}{0.1in}{-0.3in}
\topinset{(f)}{\includegraphics[scale=0.21]{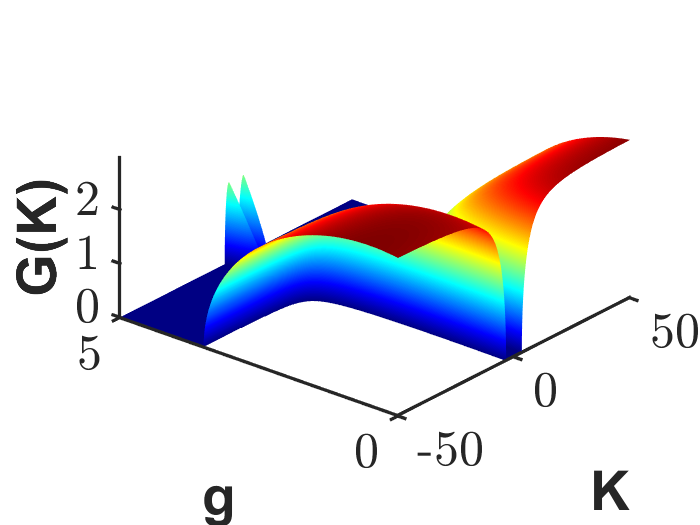}}{0.1in}{-0.3in}\\
\topinset{(g)}{\includegraphics[scale=0.21]{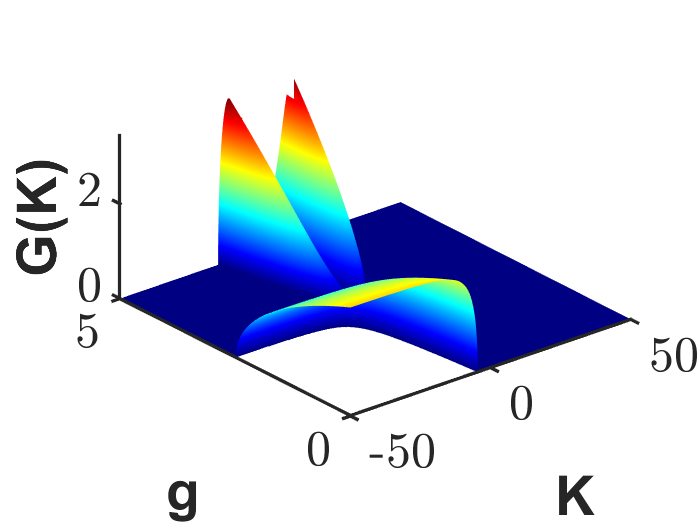}}{0.1in}{-0.3in}
\topinset{(h)}{\includegraphics[scale=0.21]{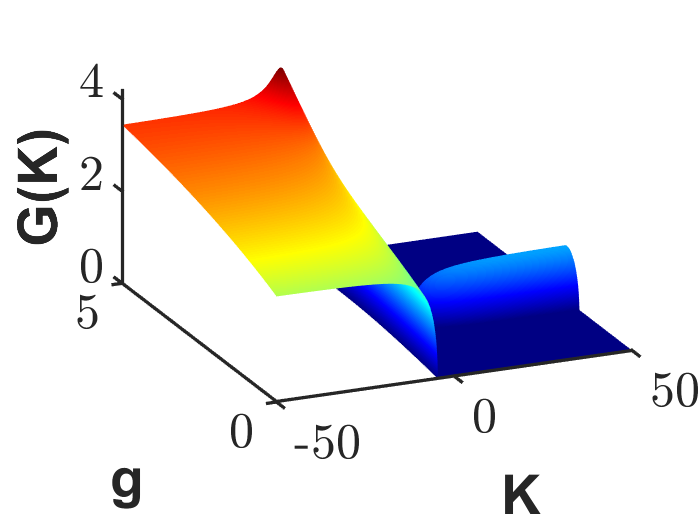}}{0.1in}{-0.3in}
\caption{(Color online) The dynamics of MI in the normal dispersion regime of the left and right incidences are shown in the left and right panels, respectively. Here, (a-b) represents the conventional case, (c-d) the unbroken $\mathcal{PT}$-symmetric, (e-f) the exceptional point, and (g-h) the broken $\mathcal{PT}$-symmetric regimes with FWM $\xi=1$, where (a-b) $f=0.1$, (c-d) $f=0.5$, (e-f) $f=1$, and (g-h) $f=3$. The other parameters are $P=1.5$, $\kappa=1$, and $\gamma=2$.}\label{Figure6}
\end{figure}
\begin{figure}[t]
   \topinset{(a)}{\includegraphics[scale=0.23]{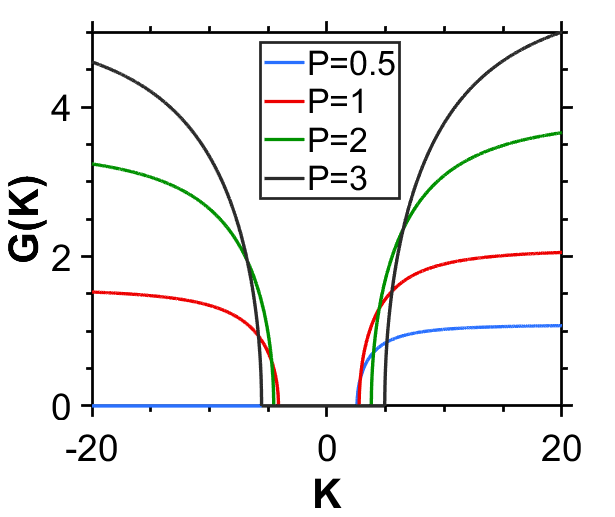}}{0.1in}{-0.3in} \topinset{(b)}{\includegraphics[scale=0.23]{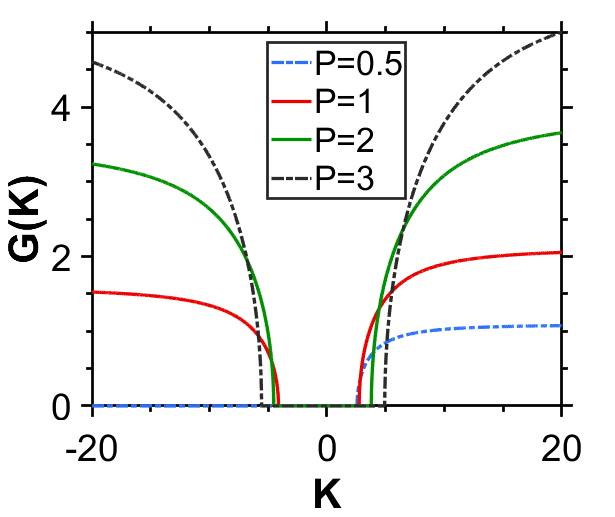}}{0.1in}{-0.3in}\\   \topinset{(c)}{\includegraphics[scale=0.23]{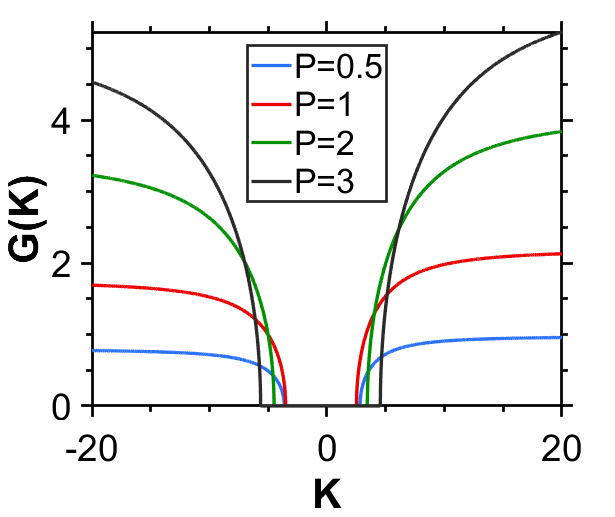}}{0.1in}{-0.3in}
 \topinset{(d)}{\includegraphics[scale=0.23]{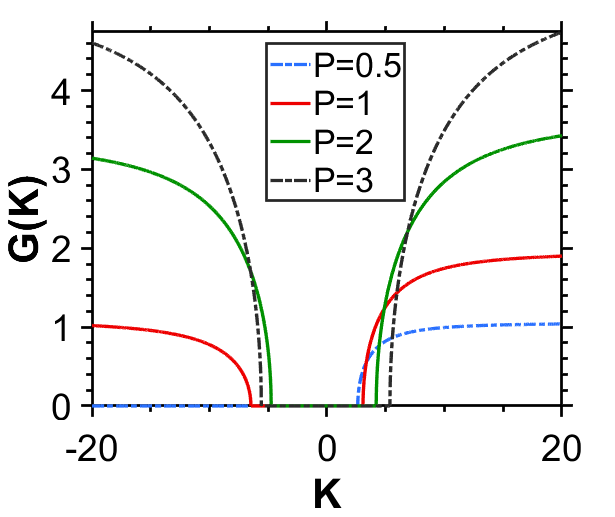}}{0.1in}{-0.3in}\\
  \topinset{(e)}{\includegraphics[scale=0.23]{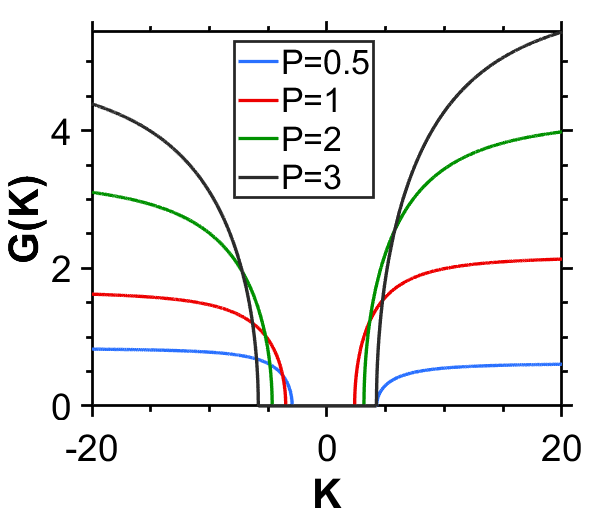}}{0.1in}{-0.3in}\topinset{(f)}{\includegraphics[scale=0.23]{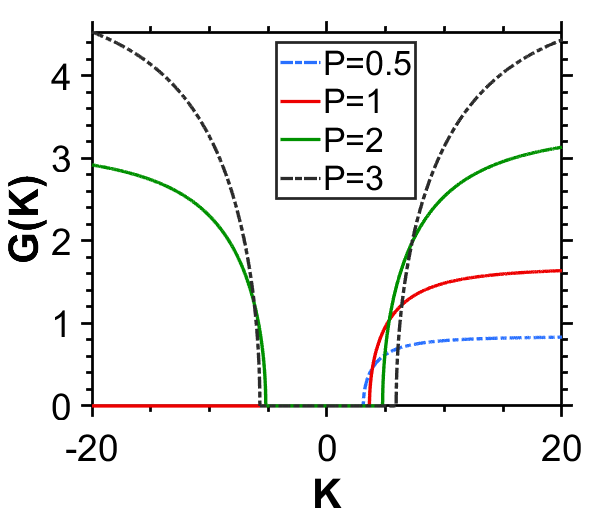}}{0.1in}{-0.3in}
 \topinset{(g)}{\includegraphics[scale=0.23]{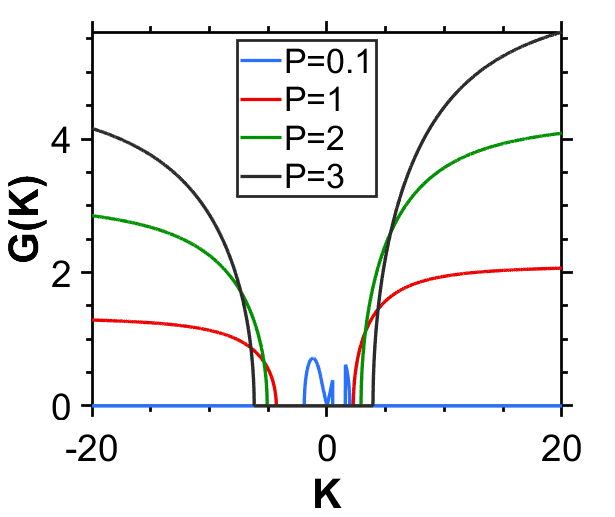}}{0.1in}{-0.3in}
 \topinset{(h)}{\includegraphics[scale=0.23]{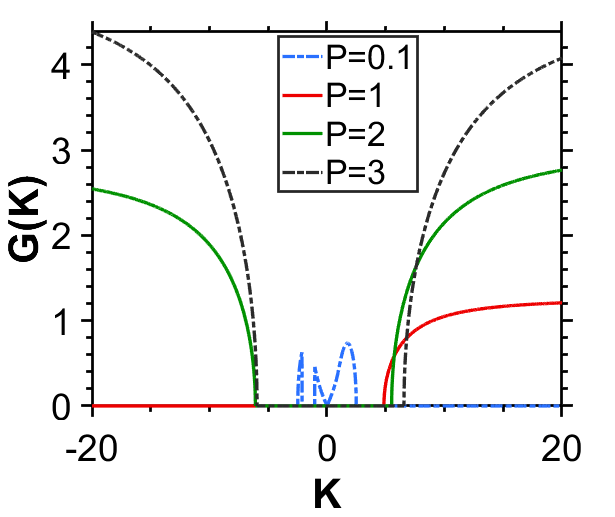}}{0.1in}{-0.3in}

\caption{(Color online) Plots show the role of power on the MI gain spectrum in the normal dispersion regime for left (left panels) and right (right panels) incidences, depicting (a-b) represent the conventional case, (c-d) the unbroken $\mathcal{PT}$-symmetric, (e-f) the exceptional point, and (g-h) the broken $\mathcal{PT}$-symmetric regimes with $f=0.8$, $\kappa=1$, $\gamma=2$, and $\xi=1$. }\label{Figure7}
\end{figure}
\begin{figure*}[t]
  \topinset{(a)}{\includegraphics[scale=0.22]{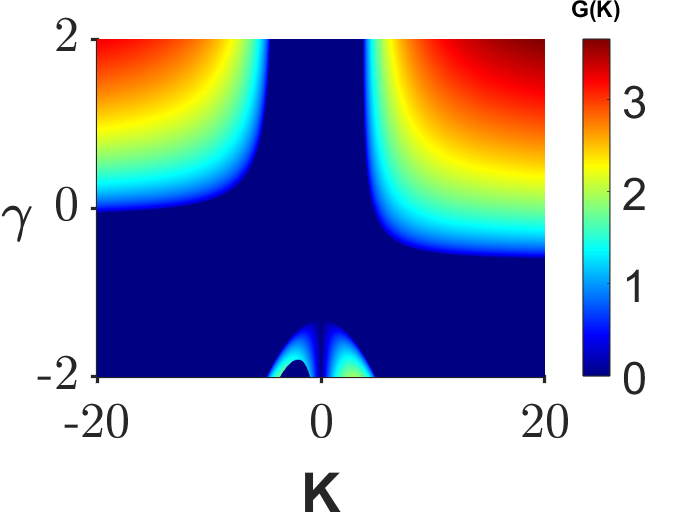}}{0.1in}{-0.3in}
 \topinset{(b)}{\includegraphics[scale=0.22]{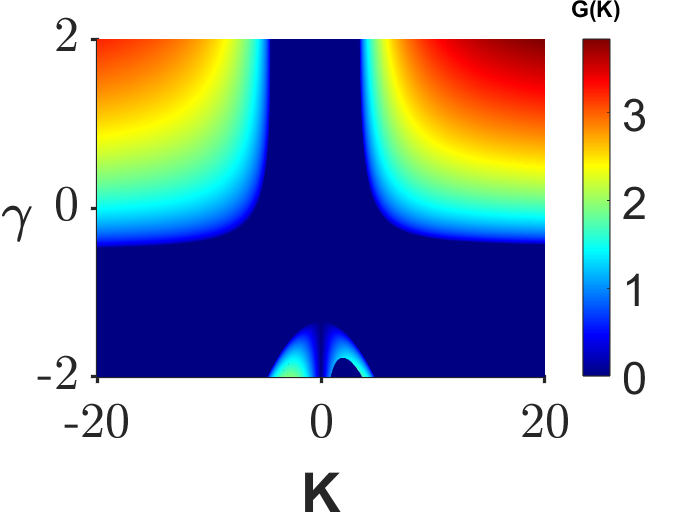}}{0.1in}{-0.3in}
 \topinset{(c)}{\includegraphics[scale=0.22]{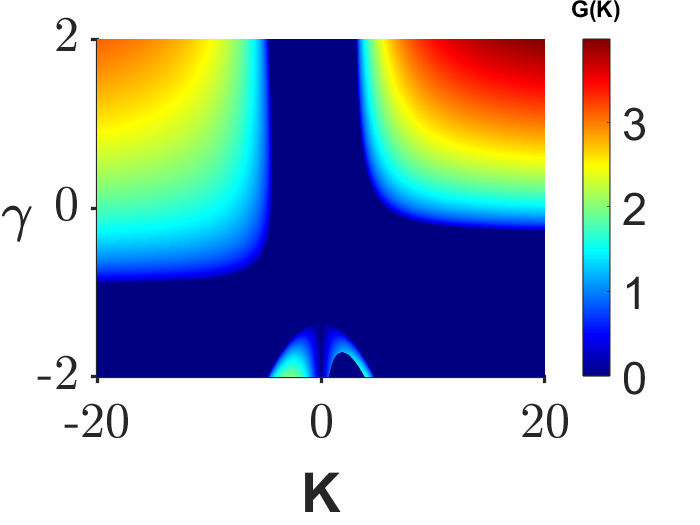}}{0.1in}{-0.3in}
 \topinset{(d)}{\includegraphics[scale=0.22]{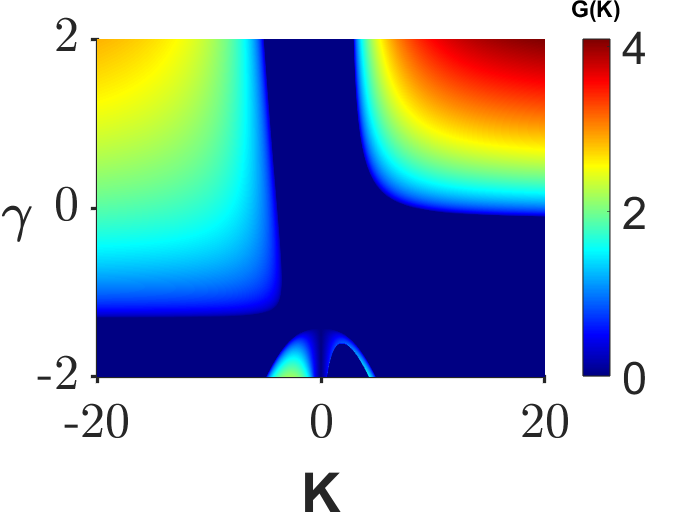}}{0.1in}{-0.3in}\\
 \topinset{(e)}{\includegraphics[scale=0.22]{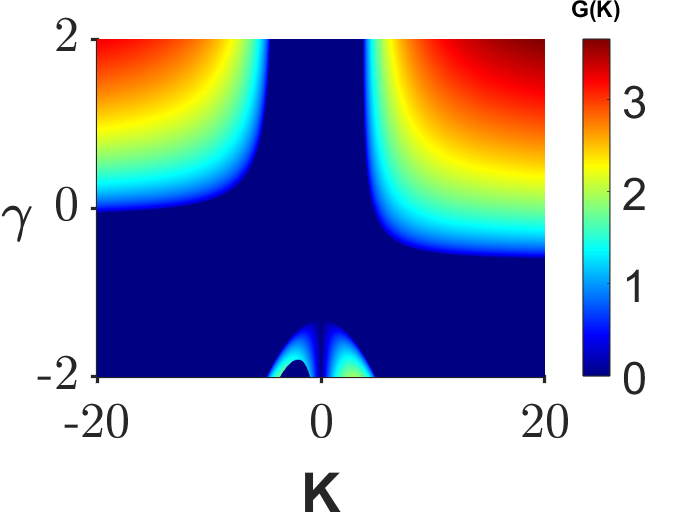}}{0.1in}{-0.3in}
\topinset{(f)}{\includegraphics[scale=0.22]{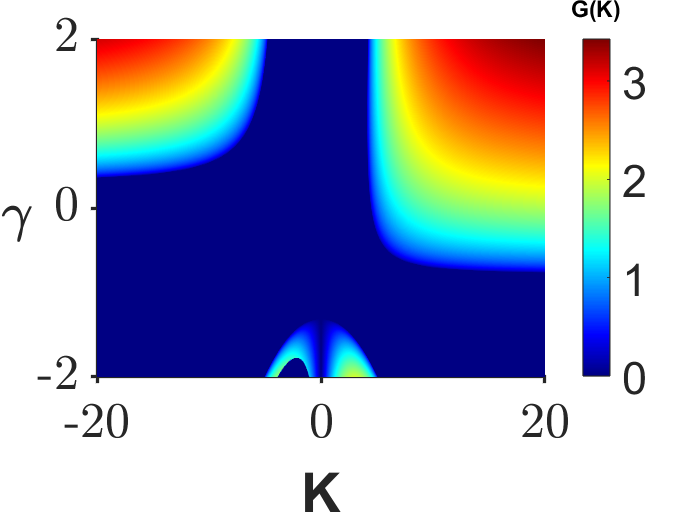}}{0.1in}{-0.3in}
\topinset{(g)}{\includegraphics[scale=0.22]{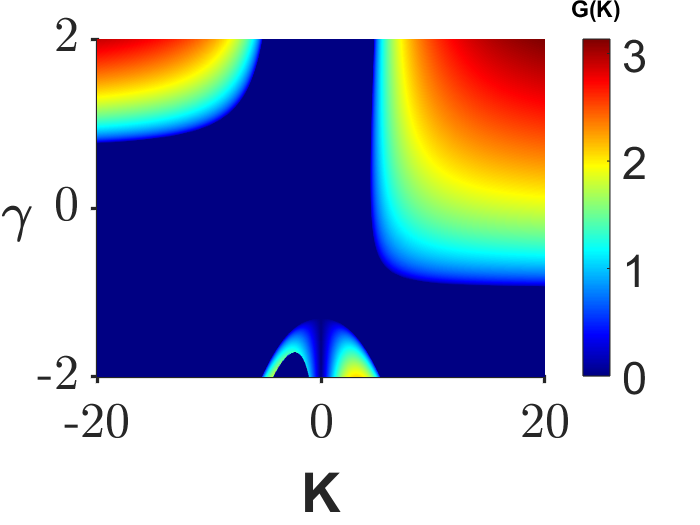}}{0.1in}{-0.3in}
\topinset{(h)}{\includegraphics[scale=0.22]{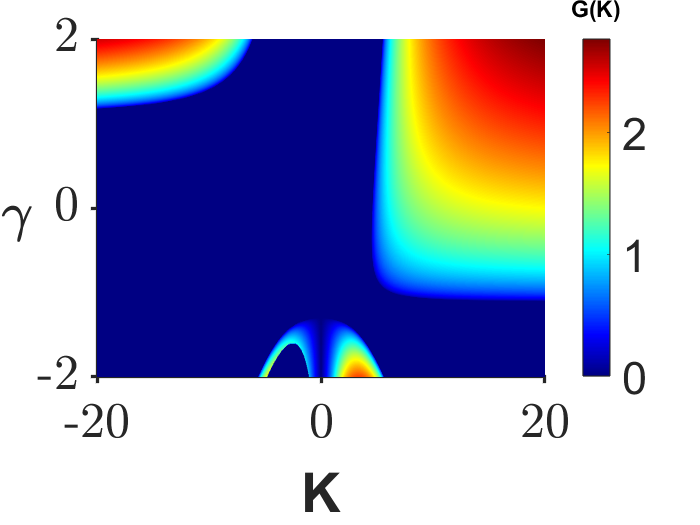}}{0.1in}{-0.3in}
\caption{(Color online) Plots illustrate the influence of the SPM effect on the MI gain spectrum in the normal dispersion regime for left (top panels) and right (bottom panels) incidences. Here (a) and (c) depict the conventional case, (b) and (f) the unbroken $\mathcal{PT}$-symmetric, (c) and (g) the exceptional point, and (d) and (h) the broken $\mathcal{PT}$-symmetric regimes with $f=0.8$,  $\kappa=1$, $P=2$, and $\xi=1$.
 }\label{Figure8}
\end{figure*}
\begin{figure*}[t]
 \topinset{(a)}{\includegraphics[scale=0.28]{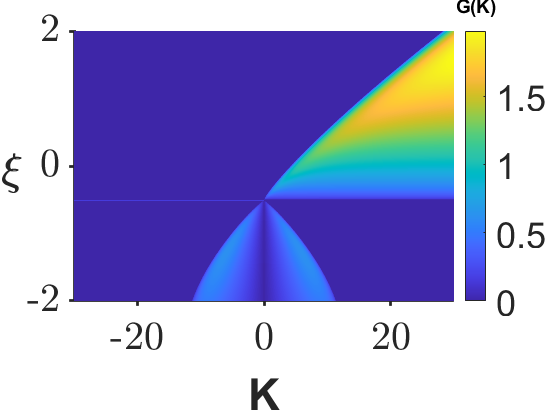}}{0.1in}{-0.3in}
\topinset{(b)}{\includegraphics[scale=0.28]{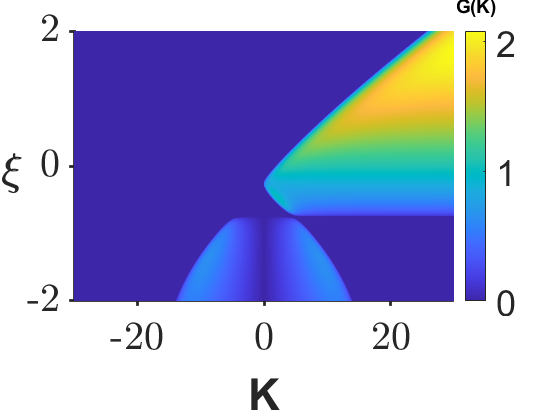}}{0.1in}{-0.3in}
\topinset{(c)}{\includegraphics[scale=0.28]{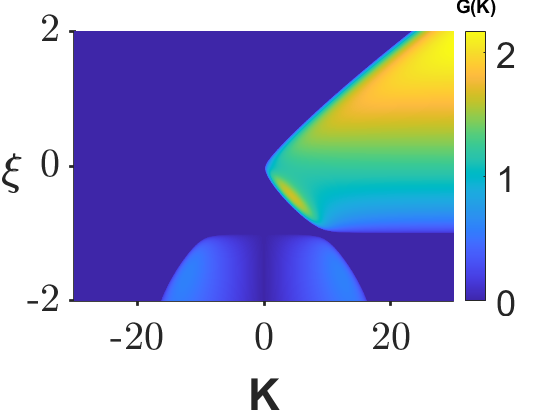}}{0.1in}{-0.3in}
\topinset{(d)}{\includegraphics[scale=0.28]{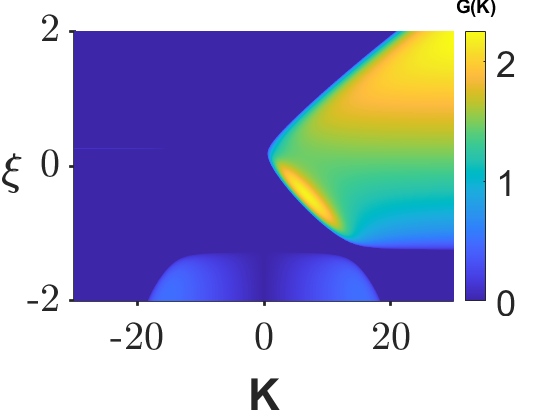}}{0.1in}{-0.3in}\\
 \topinset{(e)}{\includegraphics[scale=0.28]{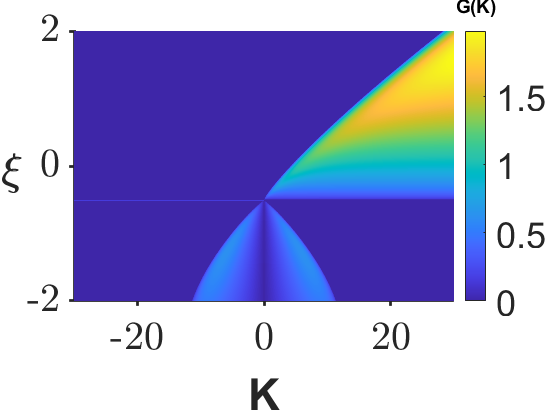}}{0.1in}{-0.3in}
\topinset{(f)}{\includegraphics[scale=0.28]{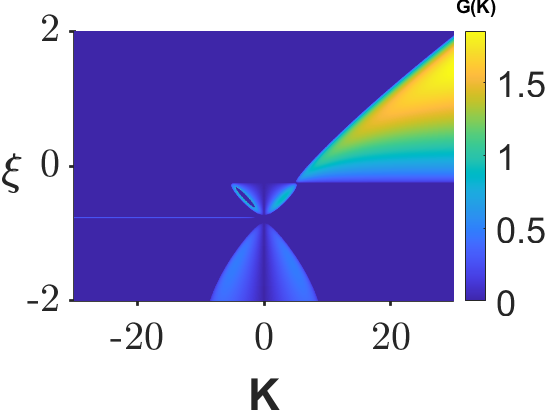}}{0.1in}{-0.3in}
\topinset{(g)}{\includegraphics[scale=0.28]{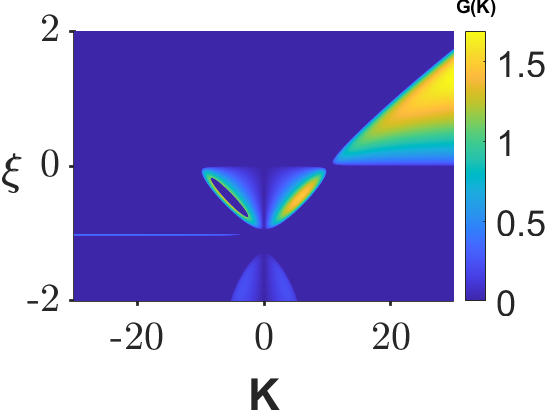}}{0.1in}{-0.3in}
\topinset{(h)}{\includegraphics[scale=0.28]{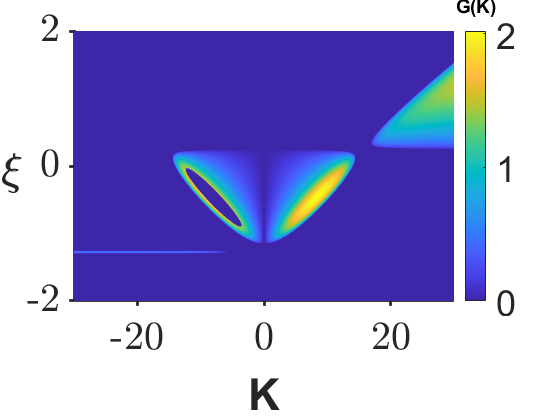}}{0.1in}{-0.3in}
\caption{(Color online) Plots illustrate the influence of FWM on the MI gain spectrum in the normal dispersion regime for left incidences (top panels) and right incidences (bottom panels). Figures (a) and (c) depict the conventional case, (b) and (f) show the unbroken $\mathcal{PT}$-symmetric, (c) and (g) illustrate the exceptional point, and panels (d) and (h) display the broken $\mathcal{PT}$-symmetric regimes, with parameters $f=0.1$, $\kappa=1$, $P=2$, and $\gamma=2$.}\label{Figure9}
\end{figure*}

\subsection{MI gain spectrum in anomalous dispersion regime}
In this subsection, we aim to explore how the proposed model exhibits the MI gain spectrum in the anomalous dispersion regime. For this purpose, we assign four different values to the parameter $f$ and set the remaining parameters to constant values as $P=1.5$,~$\kappa=1$, $\gamma=1$, and $\xi=1$. Also, in addressing the three  $\cal PT$-symmetric regimes, the value of $g$ is varied continuously. Additionally, we present the MI gain spectrum for both the left and right incidences of the proposed model, offering a detailed analysis of it. First, we examine the instability gain spectrum obtained for a high value, such as  $f=-0.1$ when the light incidence is left, shown in Fig.~\ref{Figure2}(a). Here we observe three different spectra for the continuous variation of  $g$ values. For example, for both the unbroken and $\cal PT$-symmetric threshold regimes, a typical (or primary) MI gain spectrum is obtained. As the value of $g$ increases, the bandwidth shrinks and the MI gain value decreases. Except for $g=2.4$, where a distinct narrow spectrum is evident to the left of $K=0$, the system remains stable across the range $2<g<2.6$. Even with variations in the wavenumber over a broad spectrum of input wavelengths, no sidebands are observed. Further, increasing the value of  $g$ results in two distinct MI gain spectra on either side. Specifically, there is a prominent MI gain spectrum centered around the zero wavenumber on the right side, while the left side displays both a primary and a secondary MI gain spectrum at negative wavenumber. Notably, the MI gain value and the bandwidth on the right side are more pronounced than those on the left. In the corresponding right incidence, displayed in Fig.~\ref{Figure2}(e), one observes the primary MI gain spectrum, followed by the monotonically increasing gain on the right-hand side of the $K$. As $g$ increases, here, the gain value and bandwidth of the primary MI spectrum get intensified.

Secondly, when the value of $f$ is increased slightly, i.e., $f=-0.5$, in the left incidence, one can witness that the previously obtained MI gain spectrum given in Fig.~\ref{Figure2}(a) has undergone significant changes. Specifically, the MI gain spectrum of both the broken and unbroken PT-symmetric regimes becomes superimposed. The MI gain value and bandwidth of the instability gain spectrum are significantly reduced. However, the peculiar spectrum evolves with considerable enhancement in both the gain and bandwidth. Interestingly, in the corresponding right incidence cases, it begins with the standard MI gain spectrum from the broken PT-symmetric regime and evolves into a significantly larger MI gain spectrum with increasing $g$ at positive wavenumbers. In addition, the monotonically increasing gain appears alongside the former in the broken $\cal PT$-symmetric regime [see Figs.~\ref{Figure2}(b) and \ref{Figure2}(f)]. Further in the left incidence, for $f=-1$, one can observe that the two distinct MI gain spectra transform into a single pronounced MI gain spectrum at the negative wavenumber side as $g$ increases, presented in Fig. \ref{Figure2}(c). In addition to this, the secondary MI gain spectrum appears at the positive wave number side for the entire range of $g$. In the corresponding right-incidence scenario, portrayed in Fig.~\ref{Figure2}(g), one can see a mirrored version of the image shown in \ref{Figure2}(c). 

Finally, when we give a lower value for the dispersion parameter, i.e., $f=-3$, in the left incidence, see Fig. \ref{Figure2}(d), we observe a mirror-like reflected MI spectrum similar to that in the right incidence shown in Fig. \ref{Figure2}(f). That is, the primary MI gain spectrum is followed by the monotonically increasing gain on the left side of the wavenumber. In the corresponding right incidence presented in Fig. \ref{Figure2}(h), although it begins with the standard MI gain spectrum after $g$ reaches the value of $g=2$,  it transforms into two distinct MI gain spectra, with a secondary MI gain spectrum appearing on the positive wavenumber side. As $g$ further increases, the previous spectra undergo a left-right shift, becoming more pronounced than the preceding ones. In addition, a peculiar MI gain spectrum appears on the right side of the wavenumber in the range $2<g<2.5$. 

From the above illustrations, one can observe various novel MI gain spectra in the anomalous dispersion regime across different $\cal PT$-symmetric regions. These include a symmetric and asymmetric MI sideband, a secondary MI gain spectrum, a monotonically increasing gain, and a peculiar MI gain spectrum.

\subsubsection{The impact of power on MI dynamics}

In this section, we explore how power plays its role in the MI dynamics under different $\cal PT$-symmetric regimes and the conventional case. To achieve this, we will continuously change the power while keeping the values of all other parameters constant, namely  $f=-0.2$, $\kappa=1$, $\gamma=2$, and $\xi=1$. In this anomalous dispersion regime, as the system transits from the conventional case to the exceptional point, shown in Figs.~\ref{Figure3}(a),  \ref{Figure3}(c), and \ref{Figure3}(e), it yields the same type of MI gain spectrum, i.e., the symmetric MI sidebands. A notable difference is that in the left incidence of light, the gain and bandwidth of the spectrum are slightly higher. But in the corresponding right incidence case, the MI dynamics get reversed. This scenario changes when the system reaches the broken $\cal PT$-symmetric regime, illustrated in Fig.~\ref{Figure3}(g). Initially, we observed two asymmetric MI sidebands. A secondary MI gain spectrum emerges in its positive wavenumber region, where the power value typically becomes very low ($P=0.4$). With a further increase in the power value to $P=1$, the previously asymmetric MI gain spectrum at $P=0.4$ transforms into the symmetric MI sidebands, although the gain value is lower than before. However, with additional increases in power, both the gain value and bandwidth are considerably enhanced. For its exact equivalent to the $\cal PT$-symmetric regime in its right incidence, excluding the case of the spectrum at $P=0.4$ gives a continuously enhanced MI gain spectrum compared to the previous cases [see the left panels of Fig. \ref{Figure3}]. Following this, we present the maximum gain of the MI spectrum in the positive wavenumber region (both wavenumber regions exhibit identical MI gain spectra, so we present the positive wavenumber region only here) as a function of $P$ in Figs. 4(a) and (b) for four different $\cal PT$-symmetric conditions. Needless to say, the maximum MI gain of the spectrum increases with increasing power $P$.

\subsubsection{The role of SPM in the instability gain spectrum in the anomalous dispersion regime.}
Following the impact of power on the MI gain spectrum, we aim to examine the role of SPM for both left and right incidences under the four $\cal PT$-symmetric regimes, as shown in Figs.~\ref{Figure4}(a) to \ref{Figure4}(h). Similar to how the influence of power leads to an identical shape of the MI gain spectrum as the system varies from the conventional case to the exceptional point, the SPM also produces a uniform MI gain structure. In particular, negative values of the SPM lead to a monotonically increasing gain, while positive values of the SPM lead to the symmetric MI sidebands. In this case too, for the left incidence, changing from the conventional to the broken $\cal PT$-symmetric regime results in slightly higher values for both the MI gain and bandwidth. Conversely, in the right incidence, the opposite trend is observed. Finally, in the case of the broken $\cal{PT}$-symmetric regime at the left incidence shown in Fig.~\ref{Figure4}(g), except for the case when $\gamma = 2$, asymmetric MI sidebands are observed followed by a single secondary MI gain spectrum on the positive wavenumber region. Other values of $\gamma$ (varying from $\gamma=-2$ to $2$) yield spectra similar to those obtained in the previous cases of the left incidence (see Figs. \ref{Figure4} (a), \ref{Figure4} (c), and \ref{Figure4} (e)). The right incidence of the light also exhibits the MI dynamics as observed in the left incidence, as shown in Fig. \ref{Figure4}(h). However, when $\gamma = 2$, we observe an enlarged reflected image of the MI gain spectrum observed in the left incidence, as illustrated in Fig. \ref{Figure4}(h).

\subsubsection{The impact of the FWM phenomenon in the anomalous dispersion regime }

In the conventional cases of the left and right incidences, shown in Figs.~\ref{Figure5}(a) and \ref{Figure5}(e), respectively, the negative values of FWM effect yield a monotonically increasing gain up to the value of $\xi = -0.3$. After further increment of $\xi$, it transforms into two asymmetric sidebands and eventually converges to symmetric MI sidebands as $\xi$ increases further into positive values. This scenario changes slightly when the system reaches the unbroken $\cal PT$-symmetric regime, given in Fig.~\ref{Figure5}(b). Specifically, the monotonically increasing gain shifts from the symmetric to the asymmetric state. On the other hand, beyond a certain value of $\xi$, i.e. $\xi=0.3$, the asymmetric MI sidebands transform into a mirrored reflection of the corresponding left incidence. Further examining the impact of FWM on the MI dynamics in the exceptional point and broken $\cal PT$-symmetric regime of the system, we notice two main differences from the previous cases. These include the transformation of the monotonically increasing gain into the peculiar MI gain spectra and the formation of different MI structures observed in the negative wave number region as shown in Figs.~\ref{Figure5}(c) and \ref{Figure5}(d). In the right incidence of light, portrayed in Figs.~\ref{Figure5}(e) to \ref{Figure5}(g), we obtain the mirror image of the spectrum observed in the left incidence with a few changes that are discernible in the $\cal PT$-symmetric threshold and broken regimes.

\subsection{MI dynamics in normal dispersion regime}
In the following subsection, we present some important results that highlight the impact of $f$ term on the instability dynamics of the system. We wish to note that in the present case, the MI spectrum obtained under the four different $\cal PT$-symmetric conditions is entirely different from the ones obtained in the anomalous dispersion regime. For example, when $f=0.1$ under the left incidence, as shown in Fig.~\ref{Figure6}(a), one can observe that the system primarily exhibits a monotonically increasing gain on the positive wavenumber region. Specifically, a peak gain emerges at the top of the monotonically increasing gain around $g \approx 4$, while a peculiar spectrum appears in the negative wavenumber region at $g \approx 3$. The corresponding right incidence of light is presented in Fig.~\ref{Figure6}(b), producing two different MI gain spectra, including a monotonically increasing gain up to $g\approx 2$. With a further increase in the value of $g$, one can notice asymmetric MI spectra. Notably, the gain of the sidebands in the positive wavenumber region is slightly higher than that of the sideband in the negative wavenumber region. Also, at $g=3$, the MI gain spectrum in the negative wavenumber region starts to split into two sidebands. With a slight increase in the value of $f$ term as $f=0.5$, Fig.~\ref{Figure6}(c), the previous MI gain spectra are slightly modified in both incidences. For instance, in the left incidence, the gain value of the peculiar MI gain spectrum increases, whereas the value of the monotonically increasing gain decreases. Meanwhile, the asymmetric MI gain spectra become symmetric and begin to split around $g \approx 4$ on both the positive and negative wavenumber regions in the right incidence, as shown in Fig.~\ref{Figure6}(d).

With a further increase in the value of the dispersion parameter to $f=3$, the spectra primarily display asymmetric monotonically increasing gain. Upon further increase in the value of $g$, the spectra transform into symmetric MI sidebands, eventually bifurcating into two MI sidebands on both the wavenumber regions. For $f=3$, the MI structures on the left and right incidences are similar, except for a shift between them, such as the left-to-right shape becoming a right-to-left, as shown in Figs.~\ref{Figure6}(e) and \ref{Figure6}(f). As we increase the dispersion parameter to $f=5$ in the normal dispersion regime, the MI gain spectrum resembles the structure shown in Fig.~\ref{Figure6}(d), with its corresponding right incidence similar to Fig.~\ref{Figure6}(c) [see also Figs.~\ref{Figure6}(g) and \ref{Figure6}(h)].

As in the anomalous dispersion regime, the normal dispersion regime also exhibits various novel MI spectra which are observed under different $\cal PT$-symmetric regimes.

\subsubsection{The impact of power on instability (MI) in the normal dispersion regime}
The purpose of this subsection is to explore how the MI spectrum changes its characteristics due to the impact of power in the normal dispersion regime. In Fig. \ref{Figure7}, one can observe the monotonically increasing gain in all the cases of the system. However, notable changes are evident when altering the $\cal PT$-symmetric conditions. Specifically, in its conventional case for both left and right incidences, illustrated in Figs. \ref{Figure7}(a) and  \ref{Figure7}(b), the monotonically increasing gain, which initially appears in the positive wave number region at very low power values, spreads in the other wavenumber region too with the increased value of power. At the right incidence of the system, one can observe the same scenario up to its exceptional point, see Figs.~\ref{Figure7}(b),  \ref{Figure7}(d), and \ref{Figure7}(f). However, in their equivalent broken and exceptional $\cal{PT}$-symmetric regimes in the left incidence of light, shown in Figs.~\ref{Figure7}(c) and \ref{Figure7}(e), the system can generate a monotonically increasing gain in both the regions of the wavenumber at lower power values. 
Interestingly, when the value of power is very low ($P=0.1$) in the broken $\cal PT$-symmetric regime, the left incidence of light reveals asymmetric MI gain spectra, shown in Fig.~\ref{Figure7}(g). Furthermore, a secondary MI gain spectrum appears exclusively on the positive wavenumber region when the power is very low. For the same power value, the obtained MI spectra exhibit a mirror reflection of the spectrum given in the corresponding right incidence, see Fig.~\ref{Figure7}(h).

\subsubsection{The role of SPM on MI dynamics}

The role of SPM in shown in Fig.~\ref{Figure8} when the light incidence is left under various $\cal PT$-symmetric conditions. At the lowest negative SPM values, asymmetric MI gain spectra manifest, with one being more pronounced in their gain than the other. In the negative wavenumber region, an additional secondary MI gain becomes evident. As the parameter  $\gamma$ increases, these spectra eventually merge and become symmetric MI sidebands.

From the lowest negative SPM values onward (around $\gamma\approx 0.3$), asymmetric monotonically increasing  MI gain spectra begin to emerge and persist until the highest value of  $\gamma$. In these asymmetric monotonically increasing MI gains, one side consistently appears more pronounced than the other. As the parameter $g$ varies from $0$ to $1.5$, i.e.,  shifting from conventional to broken $\cal PT$-symmetric conditions, the weaker monotonically increasing MI gain intensifies, and vice versa. Similarly, in the right incidence, portrayed in Figs.~\ref{Figure8}(e) to \ref{Figure8}(h), we observe MI gain spectra analogous to those in the left incidence. However, except for $g=0$, the MI gain spectrum appears as a mirrored image of that from the left incidence, shown in Figs.~\ref{Figure8}(a) to \ref{Figure8}(d).

\subsubsection{The impact of FWM effect }
We finally investigate how the MI gain spectra undergo changes in the normal dispersion regime, induced by FWM, in different $\cal PT$-symmetric conditions. For the left incidence of light under all $\cal PT$-symmetric conditions, illustrated in Figs.~\ref{Figure9}(a) to \ref{Figure9}(d), two types of MI structures can be observed: the symmetric MI sidebands appearing for the negative values of FWM, whereas both negative and positive values of FWM reveal the peculiar MI gain structure. However, when the system shifts from the conventional to broken $\cal PT$-symmetric conditions, the symmetric MI sideband structure is suppressed, while the peculiar MI gain located at the positive wavenumber region is enhanced. 

In the corresponding right incidence, shown in Figs.~\ref{Figure9}(e) to \ref{Figure9}(h), except for $g=0$, the MI gain spectrum reveals an additional structure that resembles a combination of two tilted cones projecting opposite sides. Additionally, a peculiar MI gain spectrum can be observed on the left wavenumber region. As the parameter $g$ varies from $0$ to $1.5$, the intensity of this two conical structure in the MI gain amplifies, while the rest of the MI gain structure gets suppressed in both gain value and bandwidth. Eventually, the symmetric MI sidebands disappear when $g$ reaches the broken $\cal PT$-symmetric regime $g=1.5$.

\section{conclusion}
In conclusion, we have theoretically investigated the formation of the MI  spectrum in $\cal PT$-symmetric FBG periodic structures including the effect of FWM. Primarily, we have demonstrated nonlinear dispersion curves that exhibit diverse structures, including unique nonlinear dispersion curves with a double loop-like structure in the upper branch which has never been previously reported in the context of Bragg gratings. Based on the analysis of the dispersion curves, we have further examined the instability in two main categories: the anomalous dispersion regime and the normal dispersion regime. It is to be noted that in both of these dispersion regimes, the system has revealed the emergence of different and unusual MI gain spectra with variations in the gain/loss parameter $g$ and as a function of wavenumber across the different $\cal PT$ regimes. In particular, these spectra range from the symmetric to asymmetric MI sidebands to a secondary MI gain spectrum, a monotonically increasing gain, and a peculiar MI gain spectrum in different $\cal PT$-symmetric regimes.

Also, the impact of input power, SPM, and FWM parameters has been investigated on the emergence of MI gain spectra in both anomalous and normal dispersion regimes. In the anomalous dispersion regime, the power, SPM, and FWM parameters have caused the system to exhibit symmetric MI, asymmetric MI sidebands, and a monotonically increasing gain. In the normal dispersion regime, the system has manifested in a new MI spectrum including a two-conical structure.
 It is important to note that all of the instability spectra obtained in this study under various dispersion regimes and $\cal PT$-symmetric conditions have qualitative differences from those obtained in the conventional Bragg grating structure. 

\section*{CRediT authorship contribution statement}

{\bf I. Inbavalli}: Writing – original draft (equal), Formal analysis (equal), Investigation (equal), Software (equal). {\bf K. Tamilselvan}: Writing – original draft (equal), Formal analysis (equal), Investigation (equal), Software (equal). {\bf A. Govindarajan}: Conceptualization, Writing – original draft (equal), Methodology,  Project administration. {\bf M. Lakshmanan}: Writing- Reviewing and Editing, Funding acquisition.\\

\section*{Acknowledgement}
KT acknowledges the Department of Science and Technology (DST) and Science and Engineering Research Board (SERB), Government of India, through a National Postdoctoral Fellowship (Grant No. PDF/2021/000167). AG was supported by University Grants Commission (UGC), Government of India, through a Dr. D. S. Kothari Postdoctoral Fellowship (Grant No. F.4-2/2006 (BSR)/PH/19-20/0025). ML is supported by a DST-SERB National Science Chair position (Grant No. NSC/2020/000029) in which AG is a Visiting Scientist.
 
\end{document}